\documentclass[journal]{IEEEtran}
\usepackage{booktabs}
\usepackage{amssymb}
\usepackage{midfloat}
\usepackage{color, cite}
\usepackage{algorithm}
\usepackage{mathrsfs}
\usepackage{graphicx}
\usepackage{amsmath,bm,amssymb}
\usepackage{multicol}
\usepackage{CJK}
\usepackage{indentfirst}
\usepackage{stfloats}
\usepackage{amsmath}
\usepackage{float}
\usepackage{multirow}
\usepackage{changepage}
\usepackage{amsfonts}
\usepackage{stfloats}
\usepackage{array}
\usepackage{titletoc}
\usepackage{algorithm}
\usepackage{algorithmic}
\usepackage{multirow}
\usepackage{xcolor}
\usepackage{subfigure}
\usepackage[justification=centering]{caption} 

\ifCLASSINFOpdf
\else
\fi
\begin{document}
%
\title{Secure Interference Exploitation Precoding in MISO Wiretap Channel: Destructive Region Redefinition with Efficient Solutions}
%
%
%

\author{Ye Fan, Xuewen Liao,
       Ang Li, \textit{Member, IEEE}, and Victor C. M. Leung, \textit{Fellow, IEEE }
\thanks{Y. Fan and X. Liao are with the Department of Information and Communication Engineering, Xi'an Jiaotong University, Xi'an, 710049, China. (e-mail:fusicha900828@stu.xjtu.edu.cn, yeplos@mail.xjtu.edu.cn).}
\thanks{A. Li is with the School of Electrical and Information Engineering, The University of Sydney, Sydney, NSW, 2006, Australia. (Email: ang.li2@sydney.edu.au)}
\thanks{V. C. M. Leung is with the Department of Electrical and Computer
Engineering, The University of British Columbia, Vancouver, BC V6T 1Z4,
Canada (e-mail: vleung@ece.ubc.ca).}
}

\maketitle

\begin{abstract}
In this paper, we focus on the physical layer security for a $K$-user multiple-input-single-output (MISO) wiretap channel in the presence of a malicious eavesdropper, where we propose several interference exploitation (IE) precoding schemes for different types of the eavesdropper. Specifically, in the case where a common eavesdropper decodes the signal directly and Eve's full channel state information (CSI) is available at the transmitter, we show that the required transmit power can be further reduced by re-designing the `destructive region' of the constellations for symbol-level precoding and re-formulating the power minimization problem. We further study the SINR balancing problems with the derived `complete destructive region' with full, statistical and no Eve's CSI, respectively, and show that the SINR balancing problem becomes non-convex with statistical or no Eve's CSI. On the other hand, in the presence of a smart eavesdropper using maximal likelihood (ML) detection, the security cannot be guaranteed with all the existing approaches. To this end, we further propose a random jamming scheme (RJS) and a random precoding scheme (RPS), respectively.
To solve the introduced convex/non-convex problems in an efficient manner, we propose an iterative algorithm for the convex ones based on the Karush-Kuhn-Tucker (KKT) conditions, and deal with the non-convex ones by resorting to Taylor expansions. Simulation results show that all proposed schemes outperform the existing works in secrecy performance, and that the proposed algorithm improves the computation efficiency significantly.
\end{abstract}


%
\IEEEpeerreviewmaketitle

\section{Introduction}
\IEEEPARstart{I}{n} 5G wireless communications, there has been an ever-growing demand for the high-speed, huge-capacity, high-efficiency, and secure communications \cite{a1}. Due to the broadcast nature of the wireless signals, wireless communications are naturally facing various security threats.
Traditionally, key-based cryptographic techniques are usually employed at the upper layers to conceal information to protect the information signals from the wiretap of the potential eavesdroppers \cite{a2}-\cite{a4}. More recently, physical layer security, as a supplementary technique, has been proposed by Wyner in \cite{1} to protect the information signals from the perspective of information theory, which utilizes the channel characteristics to design the security schemes and has then received extensive research attention \cite{a5}-\cite{a7}.

One of the most popular approaches for realizing physical layer security is the jamming scheme, which is also called artificial noise (AN) scheme. If the transmitter knows the channel state information (CSI) of the legitimate user, the jamming signals can be designed in the null space of the legitimate channel to confuse the eavesdropper without interfering the legitimate transmission. For example, in \cite{2}, the jamming signal sent by the receiver in a new channel training (CT) phase was designed for security. The authors in \cite{3} studied the secure downlink transmission scheme with the help of a cooperative jammer fighting against multiple eavesdroppers. In \cite{4}, we have proposed a jamming-rate splitting scheme to achieve more secure degrees of freedom for a $K$-user multiple-input-single-output (MISO) broadcast channel with imperfect CSI at transmitter. Apart from jamming, there are also a great amount of endeavors devoted to designing precoding schemes to improve transmission secrecy \cite{r1,r2,a8,r3,r4}.  For example, in \cite{r1}, secure precoding was devised to protect the energy harvesting network, where the precoding matrix was achieved from the secrecy rate maximization problem. The authors in \cite{r2} investigated the design of
directional hybrid digital and analog precoding for the multiuser
mmwave communication system with multiple eavesdroppers. Moreover, in \cite{r3} and \cite{r4}, the authors proposed a low-complexity algorithm to optimize the secure precoding for a simultaneous wireless information and power transfer network.

Instead of treating the jamming signals or the interference as a detrimental effect at the intended receiver, another technique to manage interference for the intended receivers in physical layer security is interference alignment. For instance, the authors in \cite{5} and \cite{6} proposed to align the jamming signal to the receiving space of the eavesdroppers in multiuser networks, which makes the attackers hard to intercept the signals even with enough number of antennas. For the intended receiver, the jamming signals are aligned to an independent dimension of the information signals, which will not influence the decoding of the intended signal. Furthermore, \cite{7} directly exploited the inter-user interference as the jamming signal to ensure security.

In the above traditional approaches, the jamming signal or the interfering  signal is always cancelled or suppressed at the intended receiver. Recently, a refreshing interference exploitation (IE) technique, which is also known as constructive interference (CI) in the literature \cite{9} and realized through symbol-level precoding (SLP)  based on the instantaneous data symbol knowledge as well as the CSI \cite{8}, has overturned the traditional viewpoint on the interference in a multiuser transmission. It suggests that the interference power can further contribute to the received useful signal power and benefit the detection at the receiver with suitable precoding. Based on this viewpoint, IE-based SLP strategies have been studied for various constellations \cite{10,11,12,13,14,15}. For instance, in \cite{10,11}, the authors introduced the concept of the constructive region and investigated the non-strict phase rotation constraints for phase shift keying (PSK) modulated signals in a multiuser MISO (MU-MISO) downlink channel to achieve better detection performance. In \cite{12,13,14}, the authors further extended CI to generic multilevel modulations, i.e, quadrature amplitude modulation (QAM), in MISO and multiple-input-multiple-output (MIMO) interference channels. \cite{15} studied a spatio-temporal faster-than-Nyquist SLP method with amplitude PSK modulations in downlink multiuser MISO channels. Besides, some related works also introduced efficient algorithms to solve the SLP problems \cite{16,17,18}. In \cite{16}, the authors derived a closed-form solution for the signal-to-interference-plus-noise ratio (SINR) maximization problem for CI precoding in the multiuser downlink network, and showed that the CI scheme performs much better than the zero-forcing (ZF) scheme; in \cite{17}, a convex optimization for SLP was presented for the sum power minimization problem in a multiuser MIMO system, and a low-latency algorithm was proposed to find a heuristic solution to the optimization problem; in \cite{18}, a simplified reformulation of the power minimization problem was derived in the multiuser MISO unicast channel, where a closed-form suboptimal SLP solution was obtained using Karush-Kuhn-Tucker (KKT) optimality conditions.

Inspired by the above SLP works, the concept of IE/CI has also been extended to the field of physical layer security, where the jamming signal is designed to be constructive to the legitimate users while destructive to the eavesdropper in \cite{19,20,21}. For example, in \cite{19}, a designed jamming scheme (DJS) was proposed for the multi-eavesdropper network, where the jamming signal is obtained by minimizing the transmit power of the source under the cases with full, statistical, and no knowledge of the wiretap channels based on SLP. It is shown that the DJS scheme yields superior performance over conventional AN schemes. In \cite{20}, the authors employed the concept of the directional modulation, and proposed a non-jamming scheme (NJS) to enhance the security of multi-receivers in MIMO wiretap network in the presence of one eavesdropper. In addition, the joint physical layer security and SLP scheme has also been extended to the energy harvesting scheme in \cite{21}, where the authors demonstrated that the CI scheme yields huge power savings over traditional non-SLP schemes.

Although CI strategy was considered in the above existing works to improve the power efficiency and security performance, there exist some issues to be further addressed in terms of security. Note that in \cite{19}, the proposed IE approach only exploits part of the `complete destructive region' of the received signals at Eve, and therefore the resulting performance in \cite{19} is sub-optimal, which will be further elaborated mathematically in the following. Besides, the optimization problem in \cite{19} only considered the effect of the precoders on the power, while ignored the effect of the bound on the destructive region, and the analysis on the SINR-balancing problem was also missing. Moreover, the security of the above existing SLP schemes is realized based on the assumption that the eavesdropper decodes the signal with minimum mean-squared error (MMSE) estimation, zero-forcing (ZF), or directly decodes the symbol as the legitimate user, which is known as a common eavesdropper \cite{20}. When the eavesdropper is smart enough to utilize the maximum likelihood (ML) approach to intercept signal with known precoding strategy at the transmitter and the global CSI, the security of the schemes in \cite{19} and \cite{20} cannot be guaranteed. Therefore, in order to address the above remaining issues, we aim to guarantee the security of a $K$-user MISO wiretap channel based on IE precoding in the presence of a general eavesdropper, which can be either a common one or a smart one. For clarity, the contributions of this paper are summarized as follows:
\begin{itemize}
\item
Compared to \cite{19} and \cite{20}, for the scenario with a common eavesdropper, we re-design the CI condition for physical layer security by further including two sub-destructive regions for the case with full Eve's CSI. The corresponding power minimization problem is re-formulated by taking into account the derived `complete destructive region' and the effect of the wiretap SINR, which are jointly optimized.
\item
In addition to the power minimization problem, we further study the SINR balancing problem with full, statistical, and no Eve's CSI at the transmitter, respectively. Both the SINR thresholds for the legitimate receivers and the eavesdropper are optimized in the formulated problems to achieve a better secrecy performance.
\item
For the network with a smart eavesdropper that performs the ML detection, we propose a random jamming scheme (RJS) and a random precoding scheme (RPS), where the jamming signal and the interfering signals are designed based on ZF and strict-CI conditions, respectively.
\item
To solve the introduced convex/non-convex problems in an efficient manner, we present an iterative algorithm to solve the convex SLP problems based on the KKT conditions and penalty function, where a closed-form solution is obtained within each iteration. For the non-convex optimization problems resulting from statistical or no Eves' CSI, we utilize the Taylor expansion to deal with the non-convex constraints. The complexity of each optimization problem is also analyzed.
\end{itemize}

Simulation results validate the superiority of the proposed schemes on the security performance and the computation efficiency of the proposed algorithm. More specifically, it is shown that when the precoders and the threshold of SINR at Eve are jointly optimized in the `complete destructive region', the transmit power at the transmitter can be further reduced. For the network with a smart eavesdropper, the proposed RJS and RPS significantly improve the secrecy performance compared with the DJS \cite{19} and the NJS in \cite{20}. It is also observed that the proposed iterative algorithm outperforms CVX-based solutions markedly in terms of computation efficiency.

The remainder of this paper is organized as follows. In Section II, we introduce the MISO wiretap model and the jamming scheme. In Section III, we analyze the interference exploitation scheme for the system with a common eavesdropper, and in Section IV, we propose two random schemes for the network with a smart eavesdropper. Section V proposes an efficient algorithm for both the convex SLP problems in Section III and the non-convex SLP problems in Section IV. Simulation results are presented in Section VI and Section VII concludes the paper.

\emph{Notations:} Throughout the paper, lowercase letters denote the scalars and bold lowercase letters represent the vectors. Matrices are denoted by bold uppercase letters. ${\mathbb{C}}$ and ${\mathbb{R}}$ denote the complex and real numbers, respectively. The operators $(\cdot)^T$, $(\cdot)^H$, $ (\cdot)^*$, $(\cdot)^\dag$ represent the transpose, conjugate transposition, conjugate, and pseudo inverse operation, respectively. $|\cdot|$  denote the absolute value of a real number or the modulus of a complex number, and $\|\cdot\|_F$ represents the Frobenius norm.
%

\section{Preliminaries}
In this section, we will first describe the system model, and then introduce the traditional physical layer security scheme based on CI.
\subsection{System Model}
Consider a $K$-user MISO wiretap channel, the source (Alice) with $N$ antennas transmits confidential symbols to $K$ single-antenna users, where the data symbol $s_k$ is drawn from a unit-norm $M$-PSK constellation for user $k$ ($\text{U}_k,k\in\{1,2,...,K\}$), i.e., $s_k=e^{j\phi_k}$. There also exists a single-antenna external eavesdropper (Eve), which is near one of the users, e.g., $\text{U}_m$, and attempts to wiretap the corresponding information symbol $s_m$. To protect the confidential symbol, a jamming symbol $v=|v|e^{j\phi_v}\sim \mathcal{CN}(0,1)$ is inserted to the transmitted signal which is
\begin{equation}\label{1}
{\bf{x}}=\sum\limits_{i=1}^K{\bf{w}}_is_i+{\bf{p}}\frac{v}{|v|},
\end{equation}
where ${\bf{w}}_i\in \mathbb{C}^N$ and ${\bf{p}}\in \mathbb{C}^N$ represent the precoding vectors for $s_i$ and $v$, respectively. In this way, the received signals at $\text{U}_k$ and Eve are respectively given by
\begin{align}\label{2}
y_k&={\bf{h}}_k^T{\bf{x}}+n_k,\\
y_e&={\bf{g}}_{e}^T{\bf{x}}+n_{e},
\end{align}
where ${\bf{h}}_k\sim \mathcal{CN}({\bf{0}},{\bf{I}}_N)\in \mathbb{C}^{N}$ and ${\bf{g}}_e\sim \mathcal{CN}({\bf{0}},{\bf{I}}_N)\in \mathbb{C}^{N}$ represent the complex Gaussian channel vectors between Alice and $\{$$\text{U}_k$, Eve$\}$, respectively. Here, Alice is assumed to know the channel state information (CSI) of the legitimate channel ${\bf{h}}_k^H$, as in \cite{19,20,21}. $n_k\sim \mathcal{CN}(0,\sigma_k^2)$ and $n_{e}\sim \mathcal{CN}(0,\sigma_e^2)$ are the additive Gaussian noise at $\text{U}_k$ and Eve, respectively.


\subsection{Review of CI in Physical Layer Security}
In this section, we briefly review the existing CI approach for physical layer security in \cite{19} and explain why it is sub-optimal. Compared to traditional methods, CI-based schemes find various benefits when applied to physical layer security. On one hand, by judiciously designing the precoding strategy with CI, all interfering signals including the jamming signal can be made constructive to the information symbol, which improves the decodability of the intended symbol at the legitimate receiver. On the other hand, when the transmitter knows the full Eve's CSI, CI-based scheme can push the wiretapped signal to the destructive region of the information symbol to further degrade the performance of the Eve. In what follows, we present the corresponding mathematical CI conditions.


To be specific, firstly we rewrite the received signal $y_k$ as
\begin{align}\label{4}
y_k=\underbrace{{\bf{h}}_k^T\left( {\sum\limits_{i=1}^K{\bf{w}}_ie^{j(\phi_i-\phi_k)}+{\bf{p}}e^{j(\phi_v-\phi_k)}} \right)}_{\lambda_k}s_k+n_k.
\end{align}
For legitimate transmission, CI-based schemes exploit the available knowledge of CSI as well as the intended symbols and jamming symbols to design the precoders, which enables the received signal at $\text{U}_k$ to lie in the constructive region of the corresponding desired symbol $s_k$, as depicted in Fig. 1, where $s_k$ is assumed in the first quadrant. $t_k$ is also treated as the distance between the constructive region and the detection thresholds. Intuitively, when the received signal is located farther away from the information symbol within the constellation boundary, i.e., in the constructive zone, the detection thresholds are increased, which improves the detection performance. Based on the geometry in Fig. 1, to achieve CI, the following condition should be satisfied at $\text{U}_k$ \cite{10}:
\begin{figure}[t]
\centering
\includegraphics[width=1.8 in, height=1.7 in]{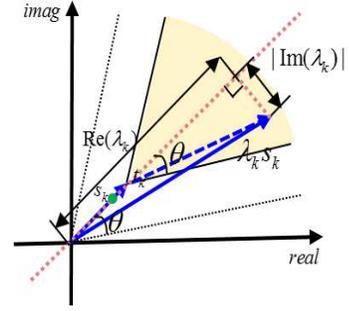}
\caption{Constructive Zone for Bob.}
\label{fig_sim}
\vspace{-0.1 in}
\end{figure}
\begin{equation}\label{5}
\frac{|\text{Im}(\lambda_k)|}{\text{Re}(\lambda_k)-t_k}\le \tan\theta,
\end{equation}
where $\theta=\frac{\pi}{M}$, and $|\text{Im}(\lambda_k)|$ and $\text{Re}(\lambda_k)$ essentially rotate the observation of the desired symbol onto the axis of the constellation symbol under consideration. Instead of being cancelled as in ZF, the jamming and interfering signals are both utilized in CI-based schemes to push the received signals farther away from the detection thresholds.$\phi$

For the wiretap transmission, we consider the case with full Eve's CSI, which is valid for the scenarios that Eve is still an active user but performs other services or has low priority compared with the legitimate users on some services in the network \cite{19}. For example, consider a video-on-demand service in a cellular network, the users who pay for this service form a group, and the others form another group. When the base station broadcasts the purchasable videos, it needs to send signals with better quality to the paid users, but avoid leaking out or just send noisy signals to the non-paid ones. If these non-paid users attempt to enjoy the service without purchase, they become the potential eavesdroppers. In this group authentication scenario, the transmitter knows the CSI of all users from the mutual communication, including the potential eavesdroppers' CSI, and the security problem under the case with full Eve's CSI is essential to be discussed.

Therefore, when Eve attempts to wiretap the information symbol $s_m,\;m\in\{1,2,...,K\}$ in the case with full Eve's CSI, the received signal $y_{e}$ in (3) is rewritten as
\begin{align}\label{6}
y_{e}=\underbrace{{\bf{g}}_{e}^T\left({\sum\limits_{i=1}^K{\bf{w}}_ie^{j(\phi_i-\phi_m)}+{\bf{p}}e^{j(\phi_v-\phi_m)}}\right)}_{\phi_e}s_m+n_{e}.
\end{align}
To avoid the interception by the eavesdropper, \cite{19} proposed the concept of `destructive region' for the case with full Eve's CSI, which is an opposite concept compared to the constructive region and aims to further distort the received signal for Eve. Geometrically, it aims to locate the wiretapped signal in the destructive regions A and B such that only incorrect data symbol can be decoded by the Eve, as shown in Fig. 2, i.e., the pink regions located in the right side of line $l_b$, which are symmetric to line $l_a$.
\begin{figure}[t]
\centering
\includegraphics[width=1.8 in, height=1.7 in]{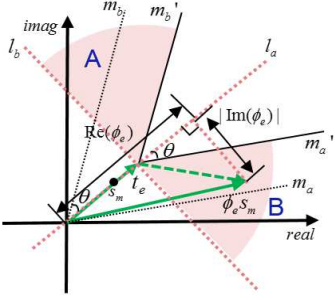}
\caption{Destructive Zone at Eve.}
\label{fig_sim}
\vspace{-0.1 in}
\end{figure}
Accordingly, the mathematical IE condition for Eve can be
\begin{align}\label{7}
&\text{Re}(\phi_e)-t_{e}\ge 0,\\
&|\text{Im}(\phi_{e})|\ge \tan\theta[\text{Re}(\phi_e)-t_{e}].
\end{align}
where $t_{e}$ is defined as the distance between the destructive zone and the detection thresholds at Eve. Based on the above analysis, it is apparent that \cite{19} does not exploit the complete destructive region and only leads to sub-optimal solutions, since the left-hand area of $l_b$ is not considered. Mathematically, this is due to the fact that \cite{19} does not consider the case when the term [Re$(\phi_e) - t_e$] becomes negative.

\section{Interference Exploitation Scheme with A Common Eavesdropper}
In this section, we focus on the IE schemes for the case with a common eavesdropper, i.e., the eavesdropper decodes the information signal intuitively without any operation as the legitimate users, where $N-K\ge 1$.  We first introduce the `complete destructive region' along with the corresponding mathematical CI conditions, based on which a power-minimization problem is proposed to optimize the precoders ${\bf{w}}$ and ${\bf{p}}$ subject to the sum transmit power constraint (\ref{5}) and the CI constraints (\ref{7})-(8) with the given $t_k$ and $t_e$.
The IE-based SINR balancing problem under the cases with full, statistical, and non Eves' CSI respectively is also studied to improve the security of the information signals at the users.
\begin{figure}[t]
\centering
\includegraphics[width=1.9 in, height=1.7 in]{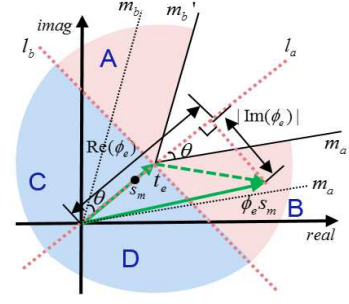}
\caption{Complete Destructive zone at Eve.}
\label{fig_sim}
\vspace{-0.1 in}
\end{figure}

\subsection{Improved Power Minimization Problem}
Compared to \cite{19} which only considers part of the destructive regions for Eve, we expand the destructive regions shown in Fig. 2 to the `complete destructive regions' in the case with full Eve's CSI, which is shown in Fig. 3. Note that in addition to the subregions A and B, the blue regions C and D located in the left side of line $l_b$ that are symmetric to line $l_a$ also denote the destructive regions, where the IE conditions for subregions C and D are given by
\begin{align}\label{9}
\text{Re}(\phi_e)-t_{e}\le 0.
\end{align}
The reason for the existence of the subregions C and D lies in the case that when the angle between the random jamming symbol and $t_e$ occasionally becomes more than $\frac{\pi}{2}$, the sum received signal will lie in C or D.
To this end, all subregions A-D formulate the `complete destructive region' for the intended symbol at Eve.


Based on the above derived complete destructive region, the power minimization problem can be constructed by jointly optimizing the SINR threshold at Eve $t_e$ and the precoders to pursue a lower power consumption at the source, given by:
\begin{subequations}\label{10}
\begin{align}
\mathcal{P}_1:\;&\mathop {\min}\limits_{{\bf{w}}_i,{\bf{p}},t_e} \;\;\left\| {\sum\limits_{i=1}^K{\bf{w}}_is_i+{\bf{p}}e^{j\phi_v}}\right\|_F^2\nonumber\\
&s.t.\;\;\;{\bf{h}}_k^T\left[{\sum\limits_{i=1}^K{\bf{w}}_ie^{j(\phi_i-\phi_k)}+{\bf{p}}e^{j(\phi_v-\phi_k)}}\right]s_k=\lambda_k s_k,\nonumber\\
&\;\;\;\;\;\;\;\;\;\;\;\;\;\;\;\;\;\;\;\;\;\;\;\;\;\;\;\;
\;\;\;\;\;\;\;\;\;\;\;\;\;\;\;\;\;\;\;\;\;\;\;\;\;\;\;\;\forall k,\\
&\;\;\;\;\;\;|\text{Im}(\lambda_k)|\le \tan\theta\left({
\text{Re}(\lambda_k)-t_k}\right),\;\;\forall k,\\
&\;\;\;\;\;\;{\bf{g}}_e^T\left({\sum\limits_{i=1}^K{\bf{w}}_ie^{j(\phi_i-\phi_m)}+{\bf{p}}e^{j(\phi_v-\phi_m)}}\right)s_m=\phi_{e} s_m,\nonumber\\
&\;\;\;\;\;\;\;\;\;\;\;\;\;\;\;\;\;\;\;\;\;\;\;\;\;\;\;\;\;\;\;\;\;\;\;\;\;\;\;m\in\{1,2,...,K\},\;\;\\
&\;\;\;\;\;\;\mathcal{X},\\
&\;\;\;\;\;\;t_e\ge 0,
\end{align}
\end{subequations}
where (10a)-(10b) denote the constructive region conditions at ${\text{U}_k}$; (10c)-(10d) denote the destructive region conditions for Eves, where $\mathcal{X}$ is the geometric conditions for subregions A-D, which are summarized in Table I for clarity. Note that since these four conditions are contradicting to each other, Alice should calculate the above optimization problem under each sub-destruction zone separately, and choose the best one, i.e., the minimum transmit power, as the final result. Overall, this optimization problem considers the effect of both the precoders and the distance between the detection threshold and destructive region on the transmit power, which is more general than the existing works in \cite{19}-\cite{20} and leads to further transmit power savings.

Note that the optimization problem $\mathcal{P}_1$ is a convex problem, and it can be readily solved by mathematical tools such as CVX. To make the proposed schemes more applicable to the practical scenarios, we will further derive a closed-form solution for the above problem with low complexity in Section V.
\begin{table}[t]
\caption{The set $\mathcal{X}$ in $\mathcal{P}_1$}
\centering
\begin{tabular}{cc}
\hline
  Subregions & $\mathcal{X}$ \\
  \hline
  A& $\text{Re}(\phi_e)-t_{e}\ge 0$,\;$\text{Im}(\phi_{e})\ge \tan\theta(\text{Re}(\phi_e)-t_{e})$\\
  B& $\text{Re}(\phi_e)-t_{e}\ge 0$,\;$\text{Im}(\phi_{e})\le -\tan\theta(\text{Re}(\phi_e)-t_{e})$  \\
  C$\&$D& $\text{Re}(\phi_e)-t_{e}\le 0$ \\
  \hline
\end{tabular}
\vspace{-0.1 in}
\end{table}
\subsection{SINR Balancing Problem}
In addition to the power minimization problem, we also consider the SINR balancing problem for the cases with full, statistical, and no Eve's CSI, respectively with the derived `complete destructive region', where both the SINR constraints at the users and the Eve have been jointly considered in the optimization problems with the precoders.

\subsubsection{Full Eve's CSI}
Consider a case with full CSI of Eve's channel at the transmitter, we attempt to maximize the minimal achievable SINR $t=\min\{t_{1},\cdots,t_{K}\}$ for all the legitimate users under the given transmit power budget, while constraining the received signal for Eve to be located in the destructive regions, which is constructed as
\begin{subequations}\label{11}
\begin{align}
\mathcal{P}_2:\;&\mathop {\min}\limits_{{\bf{w}}_i,{\bf{p}},t_e,t} \;\;-t\nonumber\\
&s.t.\;\;{\bf{h}}_k^T\left[{\sum\limits_{i=1}^K{\bf{w}}_ie^{j(\phi_i-\phi_k)}+{\bf{p}}e^{j(\phi_v-\phi_k)}}\right]s_k=\lambda_k s_k,\nonumber\\
&\;\;\;\;\;\;\;\;\;\;\;\;\;\;\;\;\;\;\;\;\;\;\;\;\;\;\;
\;\;\;\;\;\;\;\;\;\;\;\;\;\;\;\;\;\;\;\;\;\;\;\;\;\;\;\forall k,\\
&\;\;\;\;\;|\text{Im}(\lambda_k)|\le \tan\theta\left({
\text{Re}(\lambda_k)-t}\right),\;\;\forall k,\\
&\;\;\;\;\;{\bf{g}}_e^T\left({\sum\limits_{i=1}^K{\bf{w}}_ie^{j(\phi_i-\phi_m)}+{\bf{p}}e^{j(\phi_v-\phi_m)}}\right)s_m=\phi_{e} s_m,\nonumber\\
&\;\;\;\;\;\;\;\;\;\;\;\;\;\;\;\;\;\;\;\;\;\;\;\;\;\;\;\;\;\;\;\;\;\;\;\;\;\;\;m\in\{1,2,...,K\},\;\;\\
&\;\;\;\;\;\;\mathcal{X},\\
&\;\;\;\;\;\;\left\|\sum\limits_{i=1}^K{\bf{w}}_is_i+{\bf{p}}e^{j\phi_v}\right\|_F^2\le P_s,\\
&\;\;\;\;\;\;t\ge 0,\\
&\;\;\;\;\;\;t_e\ge 0,
\end{align}
\end{subequations}
where (\ref{11}e) represents the sum transmit power constraint, and $P_s$ is the maximum available transmit power at the transmitter.

\subsubsection{Statistical Eves' CSI}
When Alice can only obtain the statistical CSI of the wiretap channels by a long time observation instead of the instantaneous CSI, the wiretap ability of the Eve should be constrained with these average values. Let ${\bf{R}}_{e}=\mathbb{E}\{{\bf{g}}_e{\bf{g}}_e^H\}$ be the correlation matrix of the wiretap channel ${\bf{g}}_e$, which is assumed to be a nonsingular positive definite matrix. Then, the SINR at Eve can be expressed as
\begin{equation}\label{12}
\Gamma_{e}= \frac{{\bf{w}}_m^H{\bf{R}}_{e}{\bf{w}}_m}{\sum\limits_{i=1,i\ne m}^K{\bf{w}}_i^H{\bf{R}}_{e}{\bf{w}}_i+{\bf{p}}^H{\bf{R}}_{e}{\bf{p}}+\sigma_e^2}.
\end{equation}
Applying (\ref{12}), we can construct the statistical SINR balancing problems as
\begin{subequations}\label{13}
\begin{align}
\mathcal{P}_3:\;\;\;&\mathop {\min}\limits_{{\bf{w}}_i,{\bf{p}},t,t_e}\;\;-t\nonumber\\
&\;s.t.\;\;{\bf{h}}_k^T\left[{\sum\limits_{i=1}^K{\bf{w}}_ie^{j(\phi_i-\phi_k)}+{\bf{p}}e^{j(\phi_v-\phi_k)}}\right]s_k=\lambda_k s_k,\nonumber\\
&\;\;\;\;\;\;\;\;\;\;\;\;\;\;\;\;\;\;\;\;\;\;\;\;\;\;\;
\;\;\;\;\;\;\;\;\;\;\;\;\;\;\;\;\;\;\;\;\;\;\;\;\;\;\;\forall k,\\
&\;\;\;\;\;|\text{Im}(\lambda_k)|\le \tan\theta\left({
\text{Re}(\lambda_k)-t}\right),\;\;\forall k,\\
&\;\;\;\;\;\Gamma_e \le t_e,\\
&\;\;\;\;\;\left\|\sum\limits_{i=1}^K{\bf{w}}_is_i+{\bf{p}}e^{j\phi_v}\right\|_F^2\le P_s,\\
&\;\;\;\;\;\;t\ge 0,\\
&\;\;\;\;\;\;t_e\ge 0.
\end{align}
\end{subequations}
where the signal received at $\text{U}_k$ is still located in the constructive region of the information symbol, while for the wiretapped signal at Eve, the wiretapped SINR is constrained to be lower than $t_e$. Note that the constraint condition (\ref{13}c) denotes a non-convex set, and we will deal with it based on the Taylor expansion in Section V.

\subsubsection{No Eves' CSI}
From the above two cases with full Eves' CSI and statistical Eves' CSI, we notice that the wiretap ability of the Eve has been constrained by either the destructive region or the SINR threshold. Consider a worse case with no Eves' CSI, Alice is unable to precode the intercepted signal at Eve, and thus the CI conditions are only utilized for the useful symbols at users, i.e., we have
\begin{subequations}\label{14}
\begin{align}
\mathcal{P}_4:\;\;\;\;&\mathop {\max}\limits_{{\bf{w}}_i,{\bf{p}},t} \;\;t\nonumber\\
&\;s.t.\;\;{\bf{h}}_k^T\left[{\sum\limits_{i=1}^K{\bf{w}}_ie^{j(\phi_i-\phi_k)}+{\bf{p}}e^{j(\phi_v-\phi_k)}}\right]s_k=\lambda_k s_k,\nonumber\\
&\;\;\;\;\;\;\;\;\;\;\;\;\;\;\;\;\;\;\;\;\;\;\;\;\;\;\;
\;\;\;\;\;\;\;\;\;\;\;\;\;\;\;\;\;\;\;\;\;\;\;\;\;\;\;\forall k,\\
&\;\;\;\;\;\;|\text{Im}(\lambda_k)|\le \tan\theta\left({
\text{Re}(\lambda_k)-t}\right),\;\;\forall k,\\
&\;\;\;\;\;\;\left\|\sum\limits_{i=1}^K{\bf{w}}_is_i+{\bf{p}}e^{j\phi_v}\right\|_F^2\le P_s,\\
&\;\;\;\;\;\;\;\|{\bf{p}}\|_F^2\ge P_0,\\
&\;\;\;\;\;\;\;t\ge 0,
\end{align}
\end{subequations}
where $P_0$ is the threshold for the jamming power, and can be chosen based on the target symbol-error-rate (SER) at Eve numerically. Note that (\ref{14}d) is a non-convex constraint, and we will reformulate it into a linear constraint by Taylor expansion in Section V.

\section{Interference Exploitation Scheme with A Smart Eavesdropper}
Apart from the optimization problems in the above section which are formulated for a common eavesdropper, the eavesdropper can also be a smart one that imitates the transmission scheme at the source, which has been considered in \cite{20}. If the eavesdropper knows the modulation type and the transmission strategy at the source, it can virtually put itself in the location of the transmitter, go through $M^K$ groups of transmit symbols with the known CSI knowledge, and follow the maximum likelihood approach to find the optimal precoders. In this case, even though the eavesdropper does not know the exact jamming signal, it can still produce the equivalent precoders for the system to realize CI conditions, and the transmission scheme becomes unsafe.
To address this issue when a smart eavesdropper is present, in this section we propose two random schemes to ensure security of the network, where we consider the worst case that Alice has no knowledge of the wiretap channel.

\subsection{Random Jamming Scheme (RJS)}
For jamming schemes in the previous section, the randomness is the key factor that confuses the eavesdropper to ensure security. Thus, a natural idea is that the jamming signal ${\bf{p}}e^{j\phi_v}$ in the transmitted signal should not be treated as the constructive interfering signals, but instead be treated as the noise for all receivers. In this way, the precoder ${\bf{p}}$ should be designed in the null space of the legitimate channels so that the receivers would not be interfered. Define ${\bf{H}}=[{\bf{h}}_{1},{\bf{h}}_2,\cdots,{\bf{h}}_K]^T\in \mathbb{C}^{K \times N}$, the null space condition is denoted as ${\bf{H}}{\bf{p}}={\bf{0}}$.
Note that when ${N-K\ge 1}$, $\text{rank}({\bf{H}})=K$ holds, and the singular value decomposition (SVD) of ${\bf{H}}$ is given by ${\bf{H}}={\bf{U}}_0{\bm{\Sigma}}{\bf{V}}_0$, where the columns of ${\bf{U}}_0$ are the left singular vectors of
${\bf{H}}$, ${\bm{\Sigma}}$ is the singular values diagonal matrix, and the
columns of ${\bf{V}}_0$ are the right singular vectors of ${\bf{H}}$. Denote the last $N-K$ columns of  ${\bf{V}}_0\in \mathbb{C}^{N\times N}$ as a matrix ${\bf{V}}_1\in \mathbb{C}^{N\times (N-K)}$, the general solution of ${\bf{H}}{\bf{p}}={\bf{0}}$ can be expressed as  ${\bf{p}}=\frac{{\bf{V}}_1{\bf{k}}}{\left\|{\bf{V}}_1{\bf{k}}\right\|_F}$,
where ${\bf{k}}\in \mathbb{R}^{N-K}$ is a random vector. Hence, we rewrite the transmitted signal ${\bf{x}}$ as
\begin{equation}\label{15}
{\bf{x}}=\sum\limits_{i=1}^K{\bf{w}}_is_i+\sqrt{P_n}{\bf{p}}\frac{v}{|v|},
\end{equation}
where $P_n$ represents the allocated power for the jamming signal.
Accordingly, the SINR balancing problem that optimizes the precoding vectors and the SNR threshold for legitimate users can be formulated as
\begin{subequations}\label{16}
\begin{align}
\mathcal{P}_5:\;\;\;\;&\mathop {\max}\limits_{{\bf{w}}_i,t} \;\;t\nonumber\\
&\;\;\;s.t.
\;\;\;\;{\bf{h}}_k^T\sum\limits_{i=1}^K{\bf{w}}_ie^{j(\phi_i-\phi_k)}s_k=\tau_ks_k,
\;\;\;\forall k,\\
&\;\;\;\;\;\;\;\;\;\;\;\;|\text{Im}(\tau_k)|\le \tan\theta\left[{\text{Re}(\tau_k)-t}\right],\;\;\forall k,\\
&\;\;\;\;\;\;\;\;\;\;\;\left\|\sum\limits_{i=1}^K{\bf{w}}_is_i\right\|_F^2\le P_s-P_n,\\
&\;\;\;\;\;\;\;\;\;\;\;\;t\ge 0.
\end{align}
\end{subequations}
In this way, the received signals at $\text{U}_k$ and Eve are respectively given by
\begin{align}\label{17}
y_k&=\tau_ks_k+n_k,\\
y_e&={\bf{g}}_e^T\sum\limits_{i=1}^K{\bf{w}}_is_i+\sqrt{P_n}{\bf{g}}_e^T{\bf{p}}e^{j\phi_v}+n_e.
\end{align}
From (16)-(18), we can indicate that in the proposed random jamming scheme, CI is only exploited for the legitimate users, while the randomness of the jamming signal has been preserved at Eve for security. Besides, by controlling the value of $P_n$, there exists a tradeoff between the detection performance and the security performance of the legitimate users.

\subsection{Random Precoding Scheme (RPS)}
Note that in security-critical scenarios, the system will need a high jamming power to ensure security with the aforementioned random jamming scheme, which reduces the power allocated to the information signal and leads to inferior performance for legitimate users. In order to improve the decodability of the information signal at the legitimate receivers, we propose a random precoding scheme, which utilizes a random precoding vector based on the alignment condition in \cite{16} to ensure security.

To be specific, the transmitted signal is redesigned as
\begin{equation}\label{19}
{\bf{x}}=\sum\limits_{i=1}^K{\bf{w}}_is_i+\sqrt{P_n}{\bf{p}}.
\end{equation}
Let ${\bf{s}}=[s_1,s_2,...s_K]^T\in \mathbb{C}^{K}$, ${\bf{p}}$ is designed by the following two steps:
\begin{itemize}
\item
First, we construct an intermediate variable ${\bf{\hat p}}$, which satisfies the condition ${\bf{H}}{\bf{\hat p}}={\bf{s}}$;
\item Then, ${\bf{p}}$ is obtained by normalizing the vector ${\bf{\hat p}}$, i.e., ${\bf{p}}=\frac{{\bf{\hat p}}}{\|{\bf{\hat p}}\|}$.
\end{itemize}
When $N-K\ge 1$, the condition $\text{rank}({\bf{H}})=K<N$ holds. Similar as that in RJS scheme, ${\bf{\hat p}}$ can be denoted as ${\bf{\hat p}}={\bf{V}}_1{\bf{k}}+{\bf{r}}_0$, where ${\bf{r}}_0$ is the specific solution to the equation ${\bf{H}}{\bf{\hat p}}={\bf{s}}$ that can be achieved as ${\bf{r}}_0={\bf{H}}^\dag{\bf{s}}$. It indicates that there are infinite solutions for ${\bf{\hat p}}$ due to the random ${\bf{k}}$, thus, ${\bf{\hat p}}$ can be randomly chosen among these infinite solutions. Then, with the designed ${\bf{w}}_i$ from $\mathcal{P}_5$, the received signal at $\text{U}_k$ and Eve are respectively given by
\begin{align}\label{20}
y_k&=\tau_ks_k+\sqrt{P_n}{\bf{h}}_k^T{\bf{p}}+n_k\nonumber\\
&=\left(\tau_k+\frac{\sqrt{P_n}}{\|{\bf{\hat p}}\|}\right)s_k+n_k,\\
y_e&={\bf{g}}_e^T\sum\limits_{i=1}^K{\bf{w}}_is_i+\sqrt{P_n}{\bf{g}}_e^T{\bf{p}}+n_e.
\end{align}
Different from the random jamming scheme and the traditional precoding scheme, the proposed random precoding scheme aligns the received signal of $\text{U}_k$ to the same direction of the intended symbol, and meanwhile inserts the randomness to the wiretapped signal at Eve without causing power loss at the legitimate users, which improves the power efficiency at the source.

To summarize, when the eavesdropper decodes the signal directly, the proposed precoding designs based on ${\mathcal{P}}_1$-${\mathcal{P}}_4$ can protect the information signals; when the eavesdropper is smart enough to perform ML detection, the proposed RJS and RPS are more suitable to ensure security. In practical systems, when the decoding strategy of the eavesdropper is unpredictable at the transmitter, the proposed random schemes can be regarded as more practical and general approaches for the case with no knowledge of the eavesdropper. In the following, we will provide an efficient algorithm to solve the above optimization problems $\mathcal{P}_1-\mathcal{P}_5$.

\section{Efficient Algorithms }
In this section, we propose an efficient iterative algorithm for the convex optimization problems $\mathcal{P}_1$, $\mathcal{P}_2$, and $\mathcal{P}_5$, where the closed-form solutions of the precoders are achieved in each iteration. For the non-convex problems $\mathcal{P}_3$-$\mathcal{P}_4$, we resort to Taylor expansion, which transforms them into convex ones.

\subsection{Efficient Algorithm for $\mathcal{P}_1$, $\mathcal{P}_2$, and $\mathcal{P}_5$}
Recalling that the problems $\mathcal{P}_1$-$\mathcal{P}_2$ and $\mathcal{P}_5$ are all convex, we will explore the solution of the most complicated problem $\mathcal{P}_2$ in subregion A as an example. By following a similar procedure, the proposed algorithm can also be applied to $\mathcal{P}_1$ and $\mathcal{P}_5$ as well.

In $\mathcal{P}_2$, we first rewrite the power constraint (\ref{11}e) as \cite{16}
\begin{equation}\label{22}
\sum\limits_{i=1}^K{\bf{w}}_i^H{\bf{w}}_i+{\bf{p}}^H{\bf{p}}\le \frac{P_s}{K+1}.
\end{equation}
Applying (22), we then analyze $\mathcal{P}_2$ by Lagrangian and KKT conditions, where the Lagrangian function is given by
\begin{align}\label{23}
&{\cal {L}}_1({\bf{w}}_i,{\bf{p}},t,t_e,\delta_0,\delta_1,\delta_{2,k},\delta_{3,k},\delta_4,\kappa_{1},\kappa_{2},\kappa_{3})\nonumber\\
=&-t+\sum\limits_{k = 1}^K {\delta _{2,k}}\left[{{\bf{h}}^T_k}\left(\sum\limits_{i = 1}^K {{{\bf{w}}_i}{s_i}} +{\bf{p}}e^{j\phi_v}\right)-\lambda_k s_k\right]\nonumber\\
&+\sum\limits_{k = 1}^K {{\delta _{3,k}}} \{|{\rm{Im}}({\lambda _k})|- \tan \theta [{\rm{Re}}({\lambda _k}) - t]\}
\nonumber
\end{align}
\begin{align}
&+\delta_4\left(\sum\limits_{i=1}^K{\bf{w}}_i^H{\bf{w}}_i+{\bf{p}}^H{\bf{p}}- \frac{P_s}{K+1}\right)\nonumber\\
&+\kappa_1\left[{\bf{g}}^T_e\left(\sum\limits_{i=1}^K{\bf{w}}_is_i+{\bf{p}}e^{j\phi_v}\right)-\phi_e s_m\right]+\kappa_{2}[t_{e}-\text{Re}(\phi_e)]\nonumber\\
&+\kappa_{3}\{\tan\theta[\text{Re}(\phi_e)-t_{e}]-\text{Im}(\phi_{e})\}-\delta_0t_e-\delta_1t,
\end{align}
where $\delta_0,\delta_1,\delta_{3,k},\delta_4,\kappa_{2},\kappa_{3}$ denote the non-negative Lagrangian coefficients. Based on the Lagrangian function, the KKT conditions for the optimality of $\mathcal{P}_2$ are given by
\begin{subequations}\label{24}
\begin{align}
&\frac{\partial \mathcal{L}_1}{\partial{\bf{w}}_i}=\sum\limits_{k = 1}^K {{\delta _{2,k}}{{\bf{h}}_k^T}{s_i}} +2\delta_4{\bf{w}}_i^H+{\kappa _1}{{\bf{g}}^T_e}{s_i}={\bf{0}},\\
&\frac{\partial \mathcal{L}_1}{\partial{\bf{p}}}=\sum\limits_{k = 1}^K {\delta _{2,k}}{{\bf{h}}_k^T}e^{j\phi_v}+2\delta_4{\bf{p}}^H+\kappa_{1}{\bf{g}}^T_ee^{j\phi_v}={\bf{0}},\\
&\frac{\partial \mathcal{L}_1}{\partial t_e}=-\delta_0+{\kappa _2} - {\kappa _3}\tan \theta =0\\
&\frac{\partial \mathcal{L}_1}{\partial t}=-1-\delta_1+\sum\limits_{i = 1}^K {{\delta _{3,k}}}\tan\theta=0,\\
&{\delta _{2,k}}\left[{{\bf{h}}_k}\left(\sum\limits_{i = 1}^K {{{\bf{w}}_i}{s_i}} +{\bf{p}}e^{j\phi_v}\right)-\lambda_k s_k\right]=0,\;\;\;\;\forall k,\\
&{{\delta _{3,k}}} \{|{\rm{Im}}({\lambda _k})|- \tan \theta [{\rm{Re}}({\lambda _k}) - t]\}=0,\;\;\;\;\forall k,\\
&\delta_4\left(\sum\limits_{i=1}^K{\bf{w}}_i^H{\bf{w}}_i+{\bf{p}}^H{\bf{p}}- \frac{P_s}{K+1}\right)=0,\\
&\kappa_1\left[{\bf{g}}_e\left(\sum\limits_{i=1}^K{\bf{w}}_is_i+{\bf{p}}e^{j\phi_v}\right)-\phi_e s_m\right]=0,\\
&\kappa_{2}[t_{e}-\text{Re}(\phi_e)]=0,\\
&\kappa_{3}[\tan\theta(\text{Re}(\phi_e)-t_{e})-\text{Im}(\phi_{e})],\\
&-\delta_0t_e=0,\\
&-\delta_1t=0.
\end{align}
\end{subequations}
From (\ref{24}a) and (\ref{24}b), we can obtain the optimal precoding vectors as
\begin{align}\label{25}
{\bf{w}}_i^H&=-\frac{1}{2\delta_4}\left(\sum\limits_{k=1}^K\delta_{2,k}{\bf{h}}_k^T+\kappa_{1}{\bf{g}}_e^T\right)s_i\nonumber\\
&=(\hat {\bm \delta}{\bf{H}}+{\hat \kappa}{\bf{g}}_e^T)s_i,\\
{\bf{p}}^H&
=(\hat {\bm \delta}{\bf{H}}+{\hat \kappa}{\bf{g}}_e^T)e^{j\phi_v},
\end{align}
where $\delta_{2,k}$ and $\kappa_{1}$ are complex variables, and $ \hat {\bm \delta}=[-\frac{\delta_{2,1}}{2\delta_4},-\frac{\delta_{2,2}}{2\delta_4},...,-\frac{\delta_{2,K}}{2\delta_4}]$, ${\hat \kappa}=-\frac{\kappa_2}{2\delta_4}$. Let ${\bf{b}}=[s_1,s_2,...,s_K,e^{j\phi_v}]^T$, we further obtain
\begin{equation}\label{27}
{\bf{W}}=[{\bf{w}}_1,{\bf{w}}_2,...,{\bf{w}}_K,{\bf{p}}]=[{\bf{H}}^H\hat {\bm \delta}^H+{\bf{g}}_e^*{{\hat \kappa}}^H]{\bf{b}}^H.
\end{equation}
Besides, (\ref{11}a) and (\ref{11}c) can be respectively rewritten in matrix forms as
\begin{align}\label{28}
&{\bf{H}}{\bf{W}}{\bf{b}}=\text{diag}\{\bm \lambda\} {\bf{s}},\\
&{\bf{g}}_e^T{\bf{W}}{\bf{b}}=\phi_e s_m.
\end{align}
where $\bm \lambda=[\lambda_1,\lambda_2,...,\lambda_K]^T$.
Inserting (27) into (28) and (29), the constraint coefficients $\hat {\bm \delta}^H$ and ${{\hat \kappa}}^*$ can be derived as
\begin{align}\label{30}
\hat {\bm \delta}^H&=({\bf{H}}{\bf{H}}^H)^{-1}\left(\frac{1}{K+1}\text{diag}\{\bm \lambda\} {\bf{s}}-{\bf{H}}{\bf{g}}_e^*{{\hat \kappa}}\right),\\
{\hat \kappa}^*&=\frac{1}{(K+1)a}[ \phi_es_m-{\bf{g}}^T_e{\bf{H}}^H({\bf{H}}{\bf{H}}^H)^{-1}\text{diag}\{\bm \lambda\} {\bf{s}}],
\end{align}
where $a={{\bf{g}}^T_e}[{{\bf{I}}_N} - {{\bf{H}}^H}{({\bf{H}}{{\bf{H}}^H})^{ - 1}}{\bf{H}}]{\bf{g}}^*_e$.
With (30) and (31), ${\bf{W}}$ in (27) can be expressed as a function of the variables $\text{diag}\{\bm \lambda\}$ and $\phi_e$ in a closed form, i.e.,
\begin{equation}\label{32}
{\bf{W}}=\frac{1}{K+1}({\bf{A}}\text{diag}\{\bm \lambda\}{\bf{s}} + {\bf{C}}{\phi_e}{s_m}){\bf{b}}^H,
\end{equation}
where
\begin{align}\label{33}
{\bf{A}}&=\left\{{{\bf{I}}_N} - \frac{1}{a}[{{\bf{I}}_N} - {{\bf{H}}^H}{({\bf{H}}{{\bf{H}}^H})^{ - 1}}{\bf{H}}]{\bf{g}}^*_e{{\bf{g}}^T_e}\right\}{{\bf{H}}^H}{({\bf{H}}{{\bf{H}}^H})^{ - 1}},\\
{\bf{C}}&=\frac{1}{a}[{{\bf{I}}_N} - {{\bf{H}}^H}{({\bf{H}}{{\bf{H}}^H})^{ - 1}}{\bf{H}}]{\bf{g}}_e^*.
\end{align}
Due to that $\delta_4\ne 0$, the power constraint (\ref{11}e) can be deduced as
\begin{align}\label{35}
&\;\;\;\;\|{\bf{W}}{\bf{b}}\|_F^2\le P_s\Rightarrow{\bf{b}}^H{\bf{W}}^H{\bf{W}}{\bf{b}}\le P_s \nonumber\\
&\Rightarrow{{\bm{\lambda }}^H}\underbrace{\text{diag}\{{{\bf{s}}^H}\}{{\bf{A}}^H}{\bf{A}}\text{diag}\{{\bf{s}}\}}_{{\bf{T}}_1}{\bm{\lambda }} + {{\bm{\lambda }}^H}\underbrace{\text{diag}\{{{\bf{s}}^H}\}{{\bf{A}}^H}{\bf{C}}{s_m}}_{{\bf{T}}_2}{\phi _e} \nonumber\\
&\;\;\;+ \phi _e^*\underbrace{s_m^*{{\bf{C}}^H}{\bf{A}}\text{diag}\{{\bf{s}}\}}_{{\bf{T}}_3}{\bm{\lambda }} + \phi _e^*\underbrace{s_m^*{{\bf{C}}^H}{\bf{C}}{s_m}}_{{\bf{T}}_4}{\phi _e}\le P_s\nonumber\\
&\Rightarrow{{\bm{\lambda }}^H}{{\bf{T}}_1}{\bm{\lambda }} + {{\bm{\lambda }}^H}{{\bf{T}}_2}{\phi _e} + {\phi _e}^H{{\bf{T}}_3}{\bm{\lambda }} + {\phi _e}^H{{\bf{T}}_4}{\phi _e}\le P_s,
\end{align}
Since $\bm \lambda$ and ${\phi}$ are both complex, we expand them into their real equivalence as $\bm{\hat \lambda}=[{\rm{Re}}({\bm{\lambda }}),{\rm{Im}}({\bm{\lambda }})]^T$, $\bm{\hat \Phi}=[\text{Re}({\phi}_e),\text{Im}({\Phi}_e)]^T$, and ${\bf{\hat T}}_i=[\text{Re}({\bf{T}}_i),-\text{Im}({\bf{T}}_i);\text{Im}({\bf{T}}_i),\text{Re}({\bf{T}}_i)],i\in\{1,2,3,4\}$. Accordingly, (35) can be rewritten as
\begin{equation}\label{36}
\bm{\hat \lambda}^H{\bf{\hat T}}_1\bm{\hat \lambda}+\bm{\hat \lambda}^H{\bf{\hat T}}_2\bm{\hat \Phi}+{\bm{\hat \Phi}}^H{\bf{\hat T}}_3\bm{\hat \lambda}+{\bm{\hat \Phi}}^H{\bf{\hat T}}_4{\bm{\hat \Phi}}\le P_s.
\end{equation}
Thus, $\mathcal{P}_6$ with subregion A can be transformed into
\begin{subequations}\label{37}
\begin{align}
\mathcal{P}_6:\;\;\;\;&\mathop {\min}\limits_{\bm{\hat \lambda},\bm{\hat \Phi},t_e,t} \;\;-t\nonumber\\
&\;\;\;s.t.\;\;\;\;\;\;(\ref{36}),\nonumber\\
&\;\;\;\;\;\;\;\;\;\;\;\;\;\;\frac{\text{Im}(\lambda_k)}{\tan\theta}\le \text{Re}(\lambda_k)-t,\;\;\;\;\forall k,\\
&\;\;\;\;\;\;\;\;\;\;\;\;\;\;-\frac{\text{Im}(\lambda_k)}{\tan\theta}\le \text{Re}(\lambda_k)-t,\;\;\;\;\forall k,\\
&\;\;\;\;\;\;\;\;\;\;\;\;\;\;\;t_e-\text{Re}(\phi_k)\le 0,\\
&\;\;\;\;\;\;\;\;\;\;\;\;\;\;-\frac{\text{Im}(\phi_k)}{\tan\theta}\le t_e-\text{Re}(\phi_k),\\
&\;\;\;\;\;\;\;\;\;\;\;\;\;\;\;t\ge 0,\\
&\;\;\;\;\;\;\;\;\;\;\;\;\;\;\;t_e\ge 0.
\end{align}
\end{subequations}
For simplicity, we define
\begin{align}
&{\bf{\hat T}}_5= \left[ {\begin{array}{*{20}{c}}
{ - {{\bf{I}}_K}}&{\frac{1}{{\tan \theta }}{{\bf{I}}_K}}\\
{ - {{\bf{I}}_K}}&{ - \frac{1}{{\tan \theta }}{{\bf{I}}_K}}
\end{array}} \right],\;\;{\bf{\hat T}}_6=\left[ {\begin{array}{*{20}{c}}
{ - 1}&0\\
1&{ - \frac{1}{{\tan \theta }}}
\end{array}} \right],\nonumber\\
&{\bf{F}}_1=\left[ {\begin{array}{*{20}{c}}
{{{{\bf{\hat T}}}_1}}&{{{{\bf{\hat T}}}_2}}\\
{{{{\bf{\hat T}}}_3}}&{{{{\bf{\hat T}}}_4}}
\end{array}} \right],\;{{\bf{F}}_2} = [{{{\bf{\hat T}}}_5},{{\bf{0}}_{2K \times 2}}],\;\;{{\bf{F}}_3} = [{{\bf{0}}_{2 \times 2K}},{{{\bf{\hat T}}}_6}],\nonumber\\
&\bm \gamma=[\bm{\hat \lambda}^T,\bm{\hat \Phi}^T]^T,\;{\bf{1}}=[1,...,1]^T\in \mathbb{R}^{2K},\;{\bf{1}}_0=[-1,1]^T\nonumber,
\end{align}
then, $\mathcal{P}_6$ becomes
\begin{subequations}\label{38}
\begin{align}
\mathcal{P}_7:\;\;\;\;&\mathop {\min}\limits_{\bm \gamma,t_e,t} \;\;\;t_e\nonumber\\
&\;\;s.t.\;\;\;\;\bm \gamma^T{\bf{F}}_1\bm \gamma-P_s\le 0,\\
&\;\;\;\;\;\;\;\;\;\;\;\;{\bf{F}}_2\bm \gamma+t{\bf{1}}\le {\bf{0}}_{2K},\\
&\;\;\;\;\;\;\;\;\;\;\;\;{\bf{F}}_3\bm \gamma-t_e{\bf{1}}_0\le {\bf{0}}_{2}\\
&\;\;\;\;\;\;\;\;\;\;\;\;t\ge 0,\\
&\;\;\;\;\;\;\;\;\;\;\;\;t_e\ge 0.
\end{align}
\end{subequations}
In $\mathcal{P}_7$, the precoding vectors ${\bf{w}}_i$ and ${\bf{p}}$ to be degraded to a single vector ${\bm{\gamma}}$, and the IE constraints are reformulated into the compact form in (\ref{38}b)-(\ref{38}c). To proceed, we write the Lagrangian function $\mathcal{P}_7$ as
\begin{align}\label{39}
&\mathcal{L}_2(\bm \gamma,t_e,t,\delta_0,\delta_1,\mu_0,{\bm{\mu}}_1,{\bm{\mu}}_2)\nonumber\\
=&-t-\delta_0t_e-\delta_1t+\mu_0(\bm \gamma^T{\bf{F}}_1\bm \gamma-P_s)+{\bm{\mu}}_1^T({\bf{F}}_2\bm \gamma+t{\bf{1}})\nonumber\\
&+{\bm{\mu}}_2^T({\bf{F}}_3\bm \gamma-t_e{\bf{1}}_0),
\end{align}
where $\mu_0\ge 0$, ${\bm{\mu}}_i\ge 0,i\in\{1,2\}$. We derive the KKT conditions for $\mathcal{L}_2$ as
\begin{subequations}\label{40}
\begin{align}
&\frac{\partial \mathcal{L}_2}{\partial{\bm{\gamma}}}=2\mu_0{\bf{F}}_1\bm{\gamma}+{\bf{F}}_2^T{\bm{\mu}}_1+{\bf{F}}_3^T{\bm{\mu}}_2={\bf{0}},\\
&\frac{\partial \mathcal{L}_2}{\partial t_e}=-\delta_0-{\bm{\mu}}_2^T{\bf{1}}_0=0\\
&\frac{\partial \mathcal{L}_2}{\partial t}=-1-\delta_1+{\bm{\mu}}_1^T{\bf{1}}=0,\\
&\mu_0(\bm \gamma^T{\bf{F}}_1\bm \gamma-P_s)=0,\\
&{\bm{\mu}}_1^T({\bf{F}}_2\bm \gamma+t{\bf{1}})=0,\\
&{\bm{\mu}}_2^T({\bf{F}}_3\bm \gamma-t_e{\bf{1}}_0)=0,\\
&-\delta_0t_e=0,\\
&-\delta_1t=0,
\end{align}
\end{subequations}
where (\ref{40}a) follows that ${\bf{F}}_1^T={\bf{F}}_1$. By introducing ${\bf{F}}=[{\bf{F}}_2^T,{\bf{F}}_3^T]$ and $\bm{\mu}=[\bm{\mu}_1^T,\bm{\mu}_2^T]^T$, we can deduce the closed-form expression of $\bm{\gamma}$ as
\begin{equation}\label{41}
\bm{\gamma}=-\frac{1}{2\mu_0}{\bf{F}}_1^{-1}{\bf{F}}{\bm{\mu}}.
\end{equation}
Inserting (\ref{41}) into (\ref{40}d), the dual variable $\mu_0$ is derived as
\begin{equation}\label{42}
\mu_0=\sqrt{\frac{{\bm{\mu}}^T{\bf{F}}^T{\bf{F}}_1^{-1}{\bf{F}}\bm{\mu}}{4P_s}}.
\end{equation}
According to \cite{16}, since the optimization problem $\mathcal{P}_7$ is a convex problem, and the Slater's condition is satisfied, so that the strong duality holds. Thus, we will analyze its corresponding dual problem with (\ref{40})-(\ref{41}), which can be transformed by
\begin{align}\label{43}
\mathcal{G}&=\mathop {\max}\limits_{\bm{\mu},\delta_0,\delta_1,\mu_0}\mathop {\min}\limits_{\bm{\gamma},t_e,t}\;\;\mathcal{L}_2\nonumber\\
&=\mathop {\max}\limits_{\bm{\mu}}-\sqrt{P_s {\bm{\mu}}^T{\bf{F}}^T{\bf{F}}_1^{-1}{\bf{F}}\bm{\mu}}.
\end{align}
Let ${\bf{Q}}={\bf{F}}^T{\bf{F}}_1^{-1}{\bf{F}}$, ${\bf{f}}_1=[{\bf{0}}_{2K}^T,{\bf{1}}_0^T]^T$, and ${\bf{f}}_2=[{\bf{1}}^T,{\bf{0}}_{2}^T]^T$, the dual problem can be constructed with $\mathcal{G}$ as
\begin{subequations}\label{44}
\begin{align}
\mathcal{P}_8:\;\;\;\;&\mathop {\min}\limits_{\bm \mu} \;\;\;{\bm{\mu}}^T{\bf{Q}}{\bm{\mu}}\nonumber\\
&\;\;s.t.\;\;\;-{\bm{\mu}}^T{\bf{f}}_1\ge 0,\\
&\;\;\;\;\;\;\;\;\;\;\;\bm \mu^T{\bf{f}}_2-1\ge 0.
\end{align}
\end{subequations}
\begin{algorithm}[t]
	\caption{The iterative optimization algorithm for $\mathcal{P}_{9}$}
	\label{alg:1}
	\begin{algorithmic}[1]
\STATE{Set initial values ${\bm{\mu}}^n\ge {\bf{0}}$ and $\eta$, where $\eta$ is an extremely large positive value.}
\STATE{Calculate $(\xi_1^n,\xi_2^n)$ with ${\bm{\mu}}^n$ by using (\ref{48});}
\STATE{Calculate ${\bm{\mu}}^{n+1}$ with $(\xi_1^n,\xi_2^n)$ by using (\ref{50}) and (51);}
        \WHILE{ $|{\bm{\mu}}^{n+1}-{\bm{\mu}}^{n}|>\epsilon$}
        \STATE {Update $n=n+1$, and let ${\bm{\mu}}^{n}={\bm{\mu}}^{n+1}$;}
		\STATE {Repeat steps 2 and 3;}
        \ENDWHILE
		\STATE \textbf{Return} the optimal ${\bm{\mu}}={\bm{\mu}}^{n}$, $(\xi_1,\xi_2)=(\xi_1^n,\xi_2^n)$.
	\end{algorithmic}
\end{algorithm}
\begin{table*}[t]
\caption{Complexity Analysis of the Proposed Problems}
\centering
\begin{tabular}{r||c|c}
\hline
Algorithms & Order $n$ & Complexity \\
\hline
\hline
$\mathcal{P}_1$\;\;\;\;\;&  $\mathcal{O}((K+1)N+1)$ &\;\;\;\; A$\&$B: $\ln(1/\epsilon)\sqrt{2K+3}n[n^2+n(3+2K)+3+2K]$\\
& &\;\;\;\;\;C$\&$D: $\ln(1/\epsilon)\sqrt{2K+2}n[n^2+n(2+2K)+2+2K]$\\
\hline
$\mathcal{P}_2$\;\;\;\;\;&  $\mathcal{O}((K+1)N+2)$ &\;\;\;\; A$\&$B: $\ln(1/\epsilon)\sqrt{2K+6}n[n^2+n(4+2K)+(N^2(K+1)^2+2K+4)]$\\
& &\;\;\;\;\;C$\&$D: $\ln(1/\epsilon)\sqrt{2K+5}n[n^2+n(3+2K)+(N^2(K+1)^2+2K+3)]$\\
\hline
$\mathcal{P}_3$\;\;\;\;\;&  $\mathcal{O}((K+1)N+2)$ & $\ln(1/\epsilon)\sqrt{2K+6}n[n^2+n(2K+2)+(2K+2+N^2(K+1)^2+(N+1)^2)]$\\
\hline
$\mathcal{P}_4$\;\;\;\;\;&  $\mathcal{O}((K+1)N+1)$ & $\ln(1/\epsilon)\sqrt{2K+4}n[n^2+n(2K+2)+(2K+2+N^2(K+1)^2)]$\\
\hline
$\mathcal{P}_5$\;\;\;\;\;&  $\mathcal{O}(KN+1)$ & $\ln(1/\epsilon)\sqrt{2K+3}n[n^2+n(2K+1)+2K+1+K^2N^2]$\\
\hline
\end{tabular}
\end{table*}
To improve the calculation efficiency of the optimization problem, we propose an iterative algorithm with penalty method by reformulating $\mathcal{P}_8$ into a non-constrained problem. To be specific, we first transform (\ref{44}a)-(\ref{44}b) into equality constraints by introducing auxiliary variables ${\bf{\xi}}_1$ and ${\bf{\xi}}_2$, respectively, i.e.,
\begin{subequations}\label{45}
\begin{align}
&-{\bm{\mu}}^T{\bf{f}}_1=\xi_1\ge 0,\\
&\bm \mu^T{\bf{f}}_2-1=\xi_2\ge 0.
\end{align}
\end{subequations}
Following the principle of the penalty method, $\mathcal{P}_8$ can be transformed into the following unconstrained problem:
\begin{align}\label{46}
\mathcal{P}_{9}:&\mathop {\min}\limits_{\bm \mu,\xi_1,\xi_2} {\bm{\mu}}^T{\bf{Q}}{\bm{\mu}}+\eta[(-{\bm{\mu}}^T{\bf{f}}_1-\xi_1)^2
+(\bm \mu^T{\bf{f}}_2-1-\xi_2)^2],
\end{align}
where $\eta$ is the penalty factor. To solve $\mathcal{P}_{9}$, we propose an iterative method by alternating optimizing the variables ${\bm{\mu}}$ and $(\xi_1,\xi_2)$ as follows:
\begin{itemize}
\item First, we optimize ${\bm{\mu}}$ by setting $(\xi_1,\xi_2)$ fixed, where the objective function (\ref{46}) is denoted as $f(\bm\mu)$. Since the objective function of $\mathcal{P}_9$ is in a quadratic form, the derivative of $f_1(\bm\mu)$ with respect to $\bm\mu$ can be expressed as
    \begin{align}\label{47}
    \frac{df_1(\bm\mu)}{d \bm\mu}=&2{\bf{Q}}{\bm \mu}+2\eta[({\bf{f}}_1{\bf{f}}_1^T+{\bf{f}}_2{\bf{f}}_2^T){\bm \mu}\nonumber\\
    &+\xi_1{\bf{f}}_1-(1+\xi_2){\bf{f}}_2].
    \end{align}
    Let $\frac{df_1(\bm\mu)}{d \bm\mu}=0$, the closed-form of $\bm \mu$ is expressed as
    \begin{equation}\label{48}
    \bm\mu=\eta[{\bf{Q}}+\eta({\bf{f}}_1{\bf{f}}_1^T+{\bf{f}}_2{\bf{f}}_2^T)]^{-1}[-\xi_1{\bf{f}}_1+(1+\xi_2){\bf{f}}_2].
    \end{equation}
\item Second, we optimize $(\xi_1,\xi_2)$ by setting ${\bm{\mu}}$ fixed, where the objective function (\ref{45}) is reduced as
    \begin{align}\label{49}
    f_2(\xi_1,\xi_2)=({\bm{\mu}}^T{\bf{f}}_1+\xi_1)^2
+(\bm \mu^T{\bf{f}}_2-1-\xi_2)^2.
    \end{align}
    Calculating the partial derivative of $f_2(\xi_1,\xi_2)$ with respect to $\xi_1$ and $\xi_2$, respectively, we can obtain the optimal value of $(\xi_1,\xi_2)$ as
    \begin{align}\label{50}
    &\xi_1=-{\bm{\mu}}^T{\bf{f}}_1,\\
    &\xi_2=\bm \mu^T{\bf{f}}_2-1.
    \end{align}
\end{itemize}

Finally, we summarize the iterative algorithm in Algorithm 1. It indicates that in each iteration, a closed-form solution of ${\bm{\mu}}$ is achieved. Then, by inserting $\bm \mu$ into (\ref{41}) and combining (\ref{32}), the optimal precoding matrix can be obtained in a closed form as
\begin{align}\label{r1}
{\bf{W}}=&-\frac{1}{2\mu_0 (K+1)}({\bf{A}}\text{diag}\{{\bf{U}}_1{\bf{F}}_1^{-1}{\bf{F}}{\bm{\mu}}\}{\bf{s}} \nonumber\\
&\;\;+{\bf{u}}_2{\bf{F}}_1^{-1}{\bf{F}}{\bm{\mu}}{\bf{C}}{s_m}){\bf{b}}^H,
\end{align}
where ${\bf{U}}_1=[{\bf{I}}_K\;j{\bf{I}}_K\;{\bf{0}}_{K\times 2}]\in \mathbb{R}^{K\times 2(K+1)}$ and ${\bf{u}}_2=[{\bf{0}}_{1\times 2K}\;1\;j]\in \mathbb{R}^{1\times 2(K+1)}$ are used to transform the real-valued parameters ${\bm {\lambda}}$ and $\phi_e$ into their complex expressions.

\subsection{Linearization Algorithm for $\mathcal{P}_3$ and $\mathcal{P}_4$}
Due to the non-convexity of the constraints (\ref{15})-(\ref{16}) in $\mathcal{P}_3$ and $\mathcal{P}_4$, it is hard to solve the corresponding optimization problem directly. Besides, different from the existing algorithm in \cite{19}, $t_e$ becomes a variable to be optimized in our proposed scheme. Thus, we propose to utilize the Taylor expansion to linearize the non-convex constraints into convex ones.
\begin{figure*}
\begin{minipage}[t]{0.5\linewidth}
\centering
\includegraphics[width=3.1in, height=2.7in]{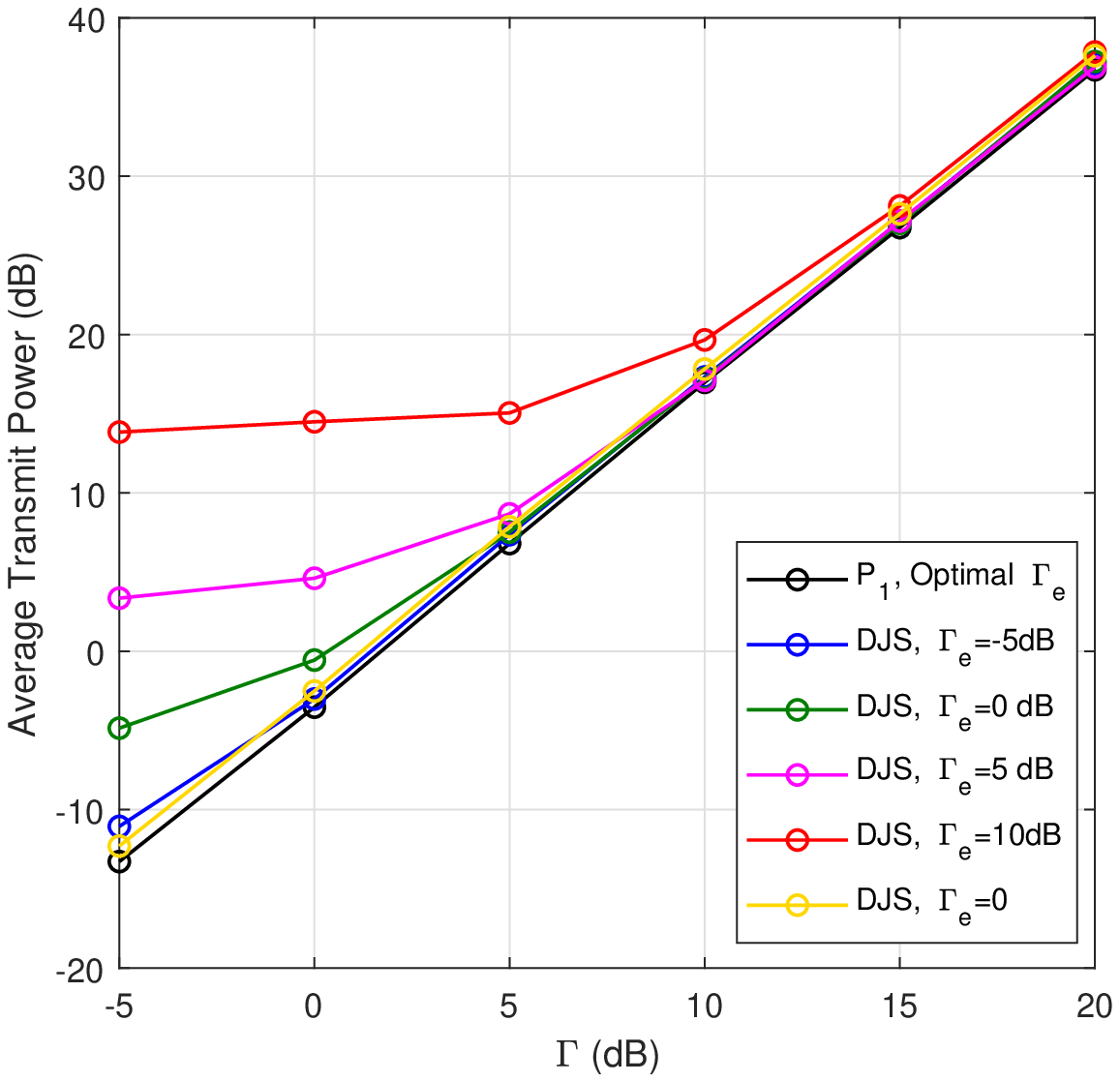}
\captionsetup{justification=centering}
\caption{Average transmit power v.s. $\Gamma$, $N=6,K=2$.}
\label{fig:side:a}
\end{minipage}
\begin{minipage}[t]{0.5\linewidth}
\centering
\includegraphics[width=3.1in, height=2.7in]{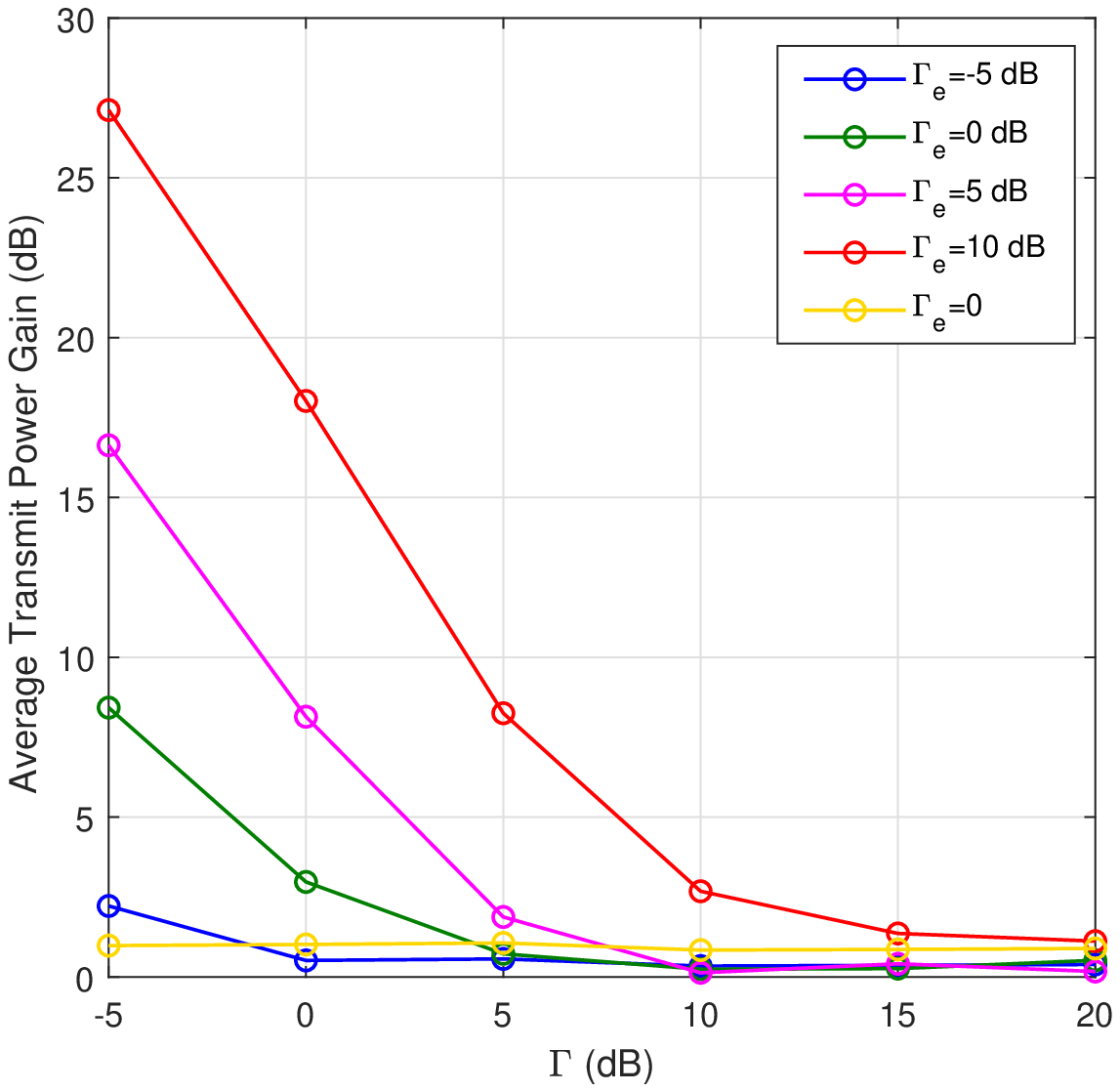}
\caption{Average transmit power gain v.s. $\Gamma$, $N=6,K=2$.}
\label{fig:side:b}
\end{minipage}
\vspace{-0.05 in}
\end{figure*}
\begin{figure*}
\begin{minipage}[t]{0.5\linewidth}
\centering
\includegraphics[width=3.1in]{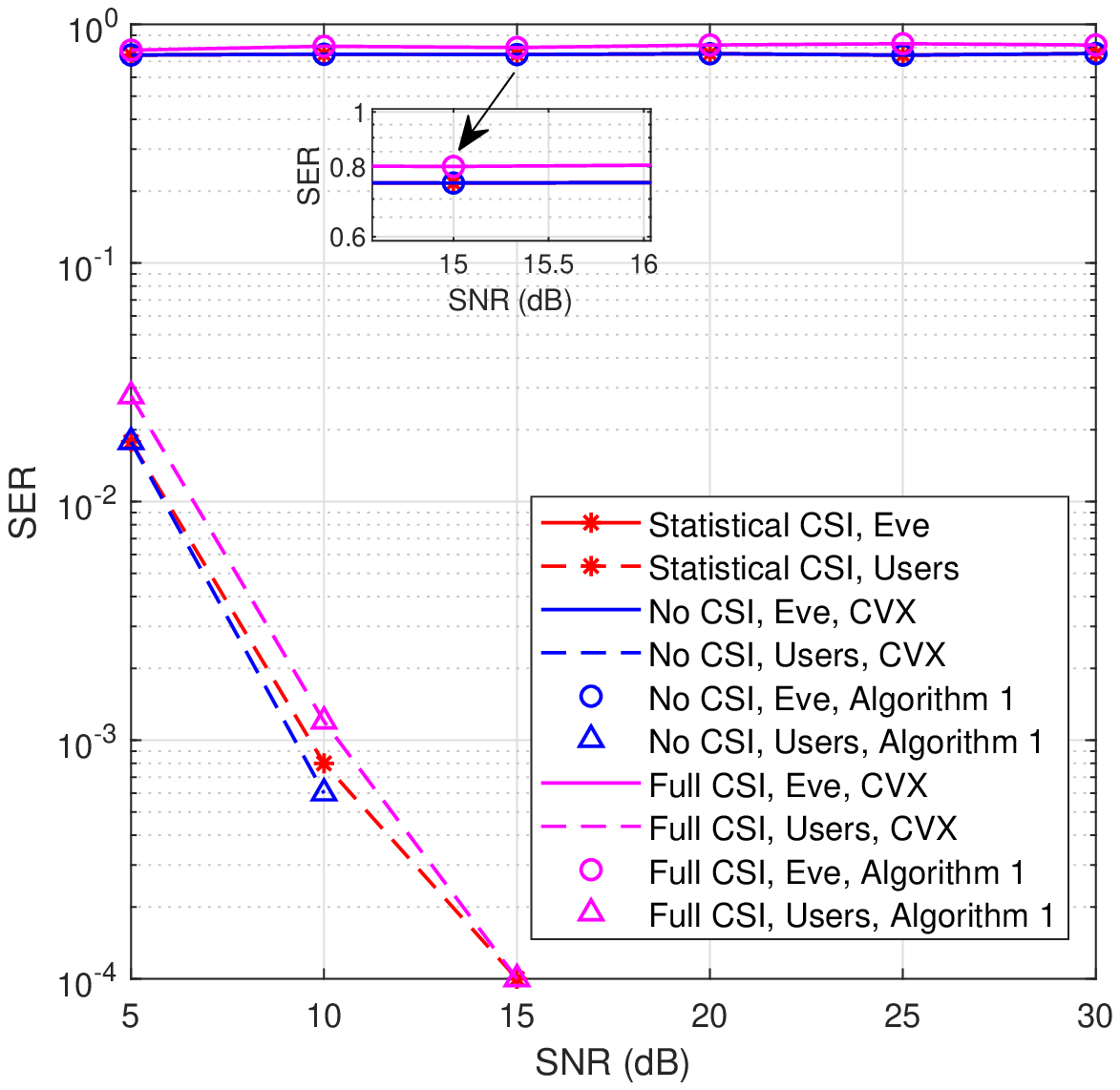}
\caption{QPSK: SER v.s. the transmit SNR, $N=6,K=2$.}
\label{fig:side:a}
\end{minipage}%
\begin{minipage}[t]{0.5\linewidth}
\centering
\includegraphics[width=3.1in]{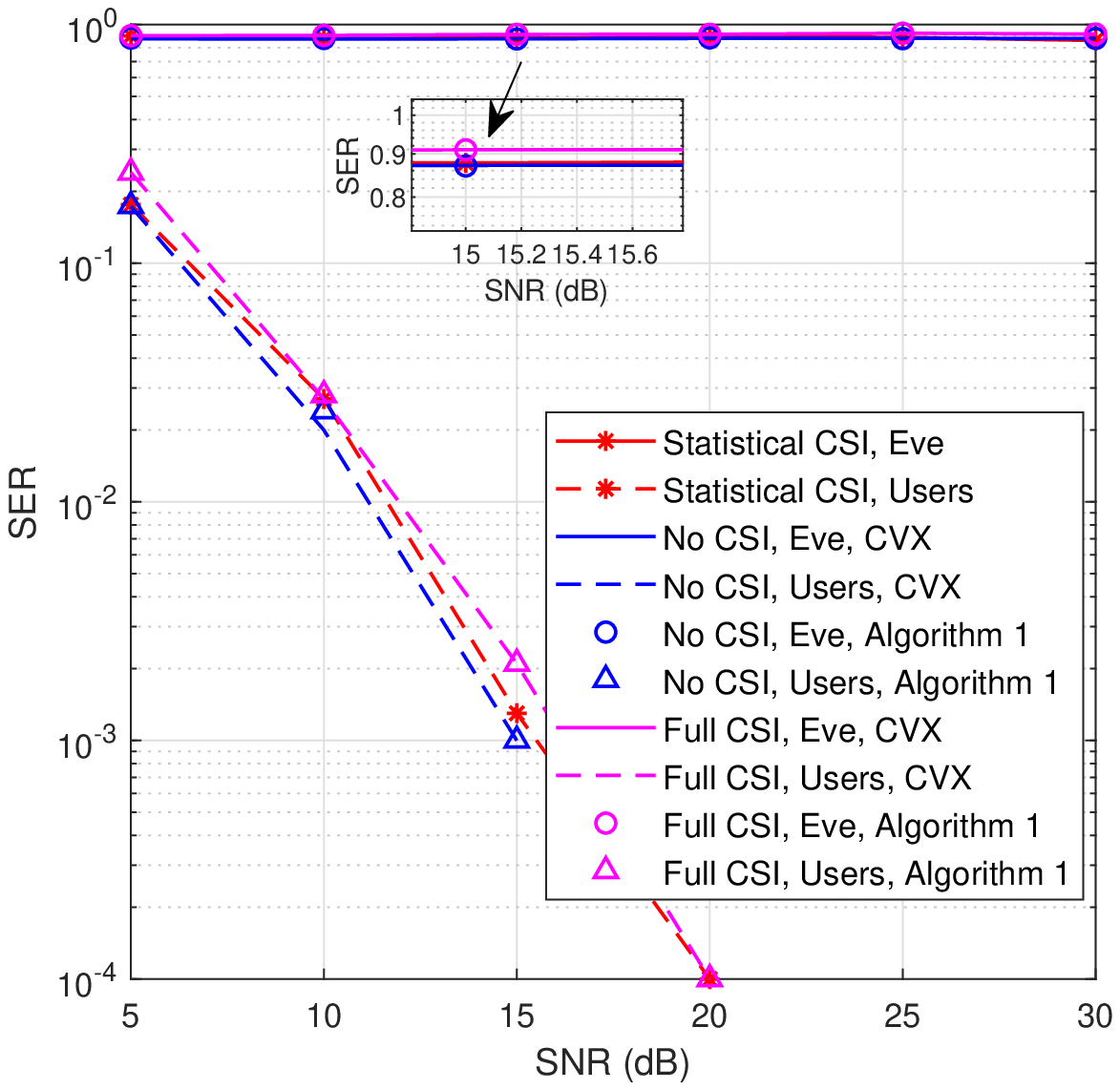}
\caption{8PSK: SER v.s. the transmit SNR, $N=6,K=2$.}
\label{fig:side:b}
\end{minipage}
\end{figure*}

Define $t_z =\frac{1}{t_e}$, the constraint (\ref{15}) becomes
\begin{equation}\label{52}
{\bf{w}}_m^H{\bf{R}}_{e}{\bf{w}}_m
\le \frac{1}{t_z}({\sum\limits_{i=1,i\ne m}^K{\bf{w}}_i^H{\bf{R}}_{e}{\bf{w}}_i+{\bf{p}}^H{\bf{R}}_{e}{\bf{p}}+\sigma_e^2}).
\end{equation}
We observe that both the left-hand side and the right-hand side of (\ref{52}) are convex, while the right-hand side is in the form of quadratic-over-linear. Thus, we attempt to linearize the right-hand side with the first-order Taylor expansion, which efficiently transforms the constraint (\ref{13}) into a convex one. To be specific, we introduce the slack variables $\tilde t_z$, $\tilde {\bf{w}}_i$, $\tilde {\bf{p}}$, and ${\tilde t}$, and perform the first-order Taylor expansion of the functions $f_{\bf{U}}({\bf{x}},y)=\frac{{\bf{x}}^H{\bf{U}}{\bf{x}}}{y}$ and $f(y)=1/y$ at the point ($\tilde {\bf{x}},\tilde y$) and $y=\tilde y$ respectively, where
\begin{equation}\label{r2}
{\mathcal{F}_{{\bf{U}}}}({\bf{x}},y,{\bf{\tilde x}},\tilde y)= \frac{{2\text{Re}({{{\bf{\tilde x}}}^H}{\bf{Ux}}) }}{{\tilde y }} - \frac{{{{{\bf{\tilde x}}}^H}{\bf{U\tilde x}}}}{{{{\tilde y }^2}}}y,
\end{equation}
\begin{equation}\label{r3}
\mathcal{F}(y,\tilde y)=\frac{1}{\tilde y}-\frac{y-\tilde y}{\tilde y^2}.
\end{equation}
To this end, (\ref{52}) can be approximated as
\begin{align}\label{53}
{\bf{w}}_m^H{\bf{R}}_{e}{\bf{w}}_m& \le \sum\limits_{i=1,i\ne m}^K \mathcal{F}_{{\bf{R}}_e}({\bf{w}}_i,t_z,{\bf{\tilde w}}_i,\tilde t_z)\nonumber\\
&\;\;\;\;+ \mathcal{F}_{{\bf{R}}_{e}}({\bf{p}},t_z,{\bf{\tilde p}},\tilde t_z)+\sigma_e^2\mathcal{F}(t_z,\tilde t_z).
\end{align}
Then, using (\ref{53}), the optimization problem $\mathcal{P}_3$ becomes a convex optimization problem, and can be effectively solved by CVX tool.


Similarly, for $\mathcal{P}_4$, the non-convex constraint (\ref{14}) is also transformed as
\begin{align}\label{54}
P_0\le \mathcal{F}({\bf{p}},{\bf{\tilde p}}),
\end{align}
where $\mathcal{F}({\bf{x}},{\bf{\tilde x}})=2\text{Re}({\bf{\tilde x}}^H{\bf{x}})-{\bf{\tilde x}}^H{\bf{\tilde x}}$ denotes the Taylor conversion of $f({\bf{x}})={\bf{x}}^H{\bf{x}}$ at the point ${\bf{x}}={\bf{\tilde x}}$. Hence, $\mathcal{P}_5$ can be solved with (\ref{54}) numerically using CVX tool.

\subsection{Computational Complexity}
Finally, we evaluate the complexity of the proposed algorithms of different types based on \cite{22}. For clarity, we summarize the computational complexity results in Table II , where $n$ denotes the number of the decision variables. The detailed analysis is shown as follows:
\begin{itemize}
\item
In $\mathcal{P}_1$, the number of the decision variables is on the order of $(K+1)N+1$.  It has $3+2K$ linear matrix inequality (LMI) constraints of size one in subregions A$\&$B, $2+2K$ LMI constraints of size one in subregions C$\&$D.
\item
In $\mathcal{P}_2$, the number of the decision variables is on the order of $(K+1)N+2$ and one SOC constraint of dimension $(K+1)N$. When the wiretapped signal is located in the subregions A$\&$B, it has $4+2K$ LMI constraints of size one; otherwise, it has $3+2K$ LMI constraints of size one.
\item
Using (\ref{53}) to replace (\ref{13}) in $\mathcal{P}_{3}$, the number of the decision variables is on the order of $(K+1)N+2$. It has $2K+2$ LMI constraints of size one, one SOC constraint of dimension $(K+1)N$, and one SOC constraint of dimensions $N+1$.
\item
Using (\ref{54}) to replace (\ref{14}) in $\mathcal{P}_{4}$, the number of the decision variables is on the order of $(K+1)N+1$. It has $2K+2$ LMI constraints of size one and one SOC constraint of dimension $(K+1)N$.
\item
In $\mathcal{P}_5$, the number of the decision variables is on the order of $KN+1$. It has $2K+2$ LMI constraints of size one, one SOC constraint of dimension $(K+1)N$, and one SOC constraint of dimension $N+1$.
\end{itemize}


\section{Numerical Results}
In this section, we will provide simulation results for the proposed  algorithms by Monte Carlo simulations. For the legitimate users, they decode the information symbol directly. For the common eavesdropper, it decodes the symbol with the same method as the legitimate users without any operation. For the smart eavesdropper, it is assumed to adopt the ML detection, and under this case, we will compare our proposed random schemes with the DJS in \cite{19} and the NJS in \cite{20}. For simplicity, we use $\Gamma_k$ and $\Gamma_e$ to denote the received SNR at user $k$ and Eve, where $t_k=\sigma_k\sqrt{\Gamma_k}$, $t_e=\sigma_e\sqrt{\Gamma_e}$ \cite{19}, and we set $\sigma_e^2=\sigma_k^2=1,\forall k$.
\begin{figure*}
\begin{minipage}[t]{0.5\linewidth}
\centering
\includegraphics[width=3.1 in]{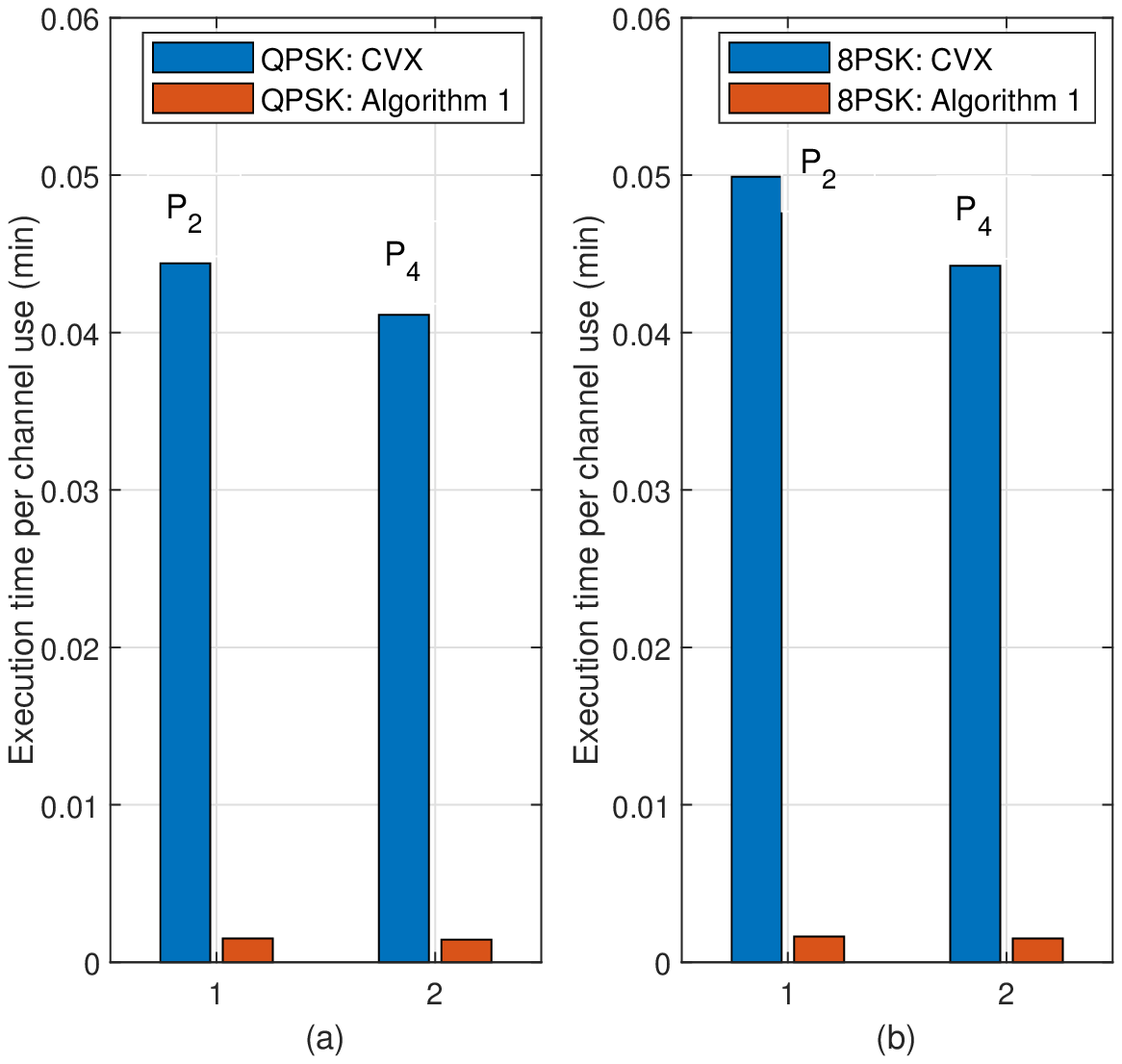}
\caption{Execution time for $\mathcal{P}_2$ and $\mathcal{P}_4$, $N=6,K=2$.}
\label{fig:side:a}
\end{minipage}
\begin{minipage}[t]{0.5\linewidth}
\centering
\includegraphics[width=3.1in, height=2.9in]{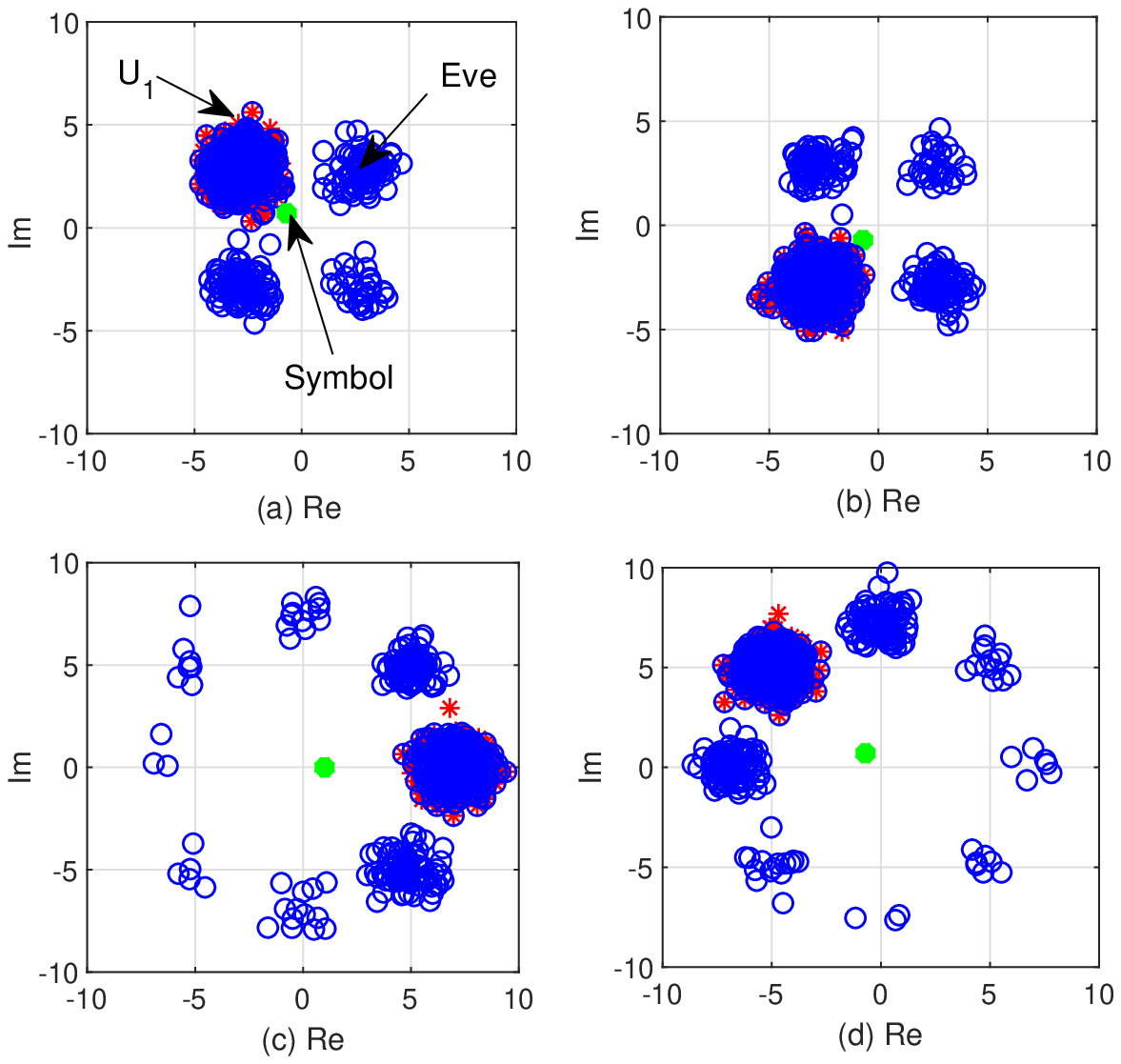}
\caption{Constellation diagrams. (a)DJS: QPSK; (b)NJS: QPSK; (C)DJS: 8PSK; (b)NJS: 8PSK. $N=6,K=2$.}
\label{fig:side:b}
\end{minipage}
\vspace{-0.05 in}
\end{figure*}
\begin{figure*}
\begin{minipage}[t]{0.5\linewidth}
\centering
\includegraphics[width=3.1in, height=2.9in]{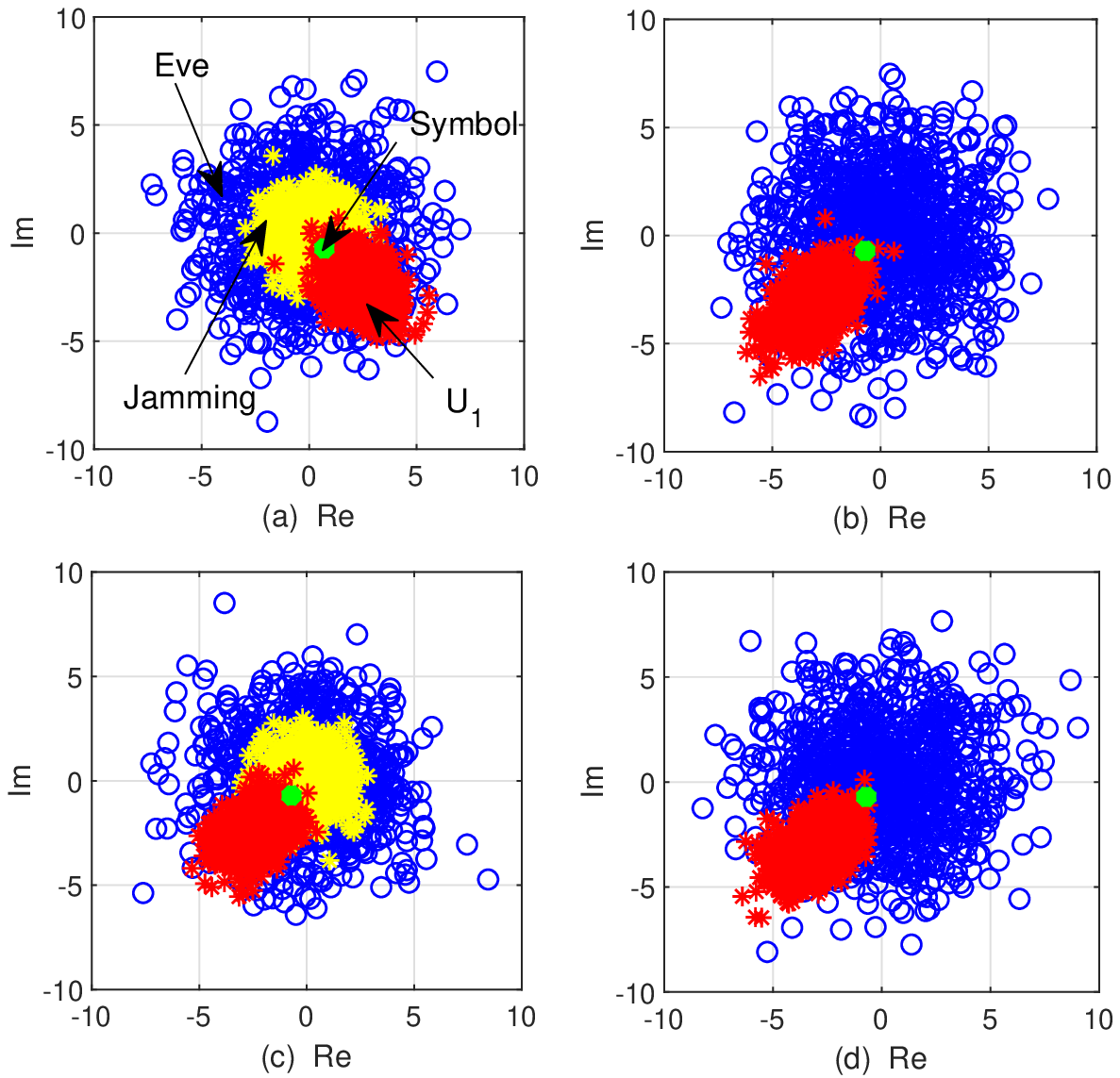}
\caption{Constellation diagrams. (a)RJS: QPSK; (b)RPS: QPSK; (C)RJS: 8PSK; (b)RPS: 8PSK. $N=6,K=2$.}
\label{fig:side:a}
\end{minipage}
\begin{minipage}[t]{0.5\linewidth}
\centering
\includegraphics[width=3.1in, height=2.9in]{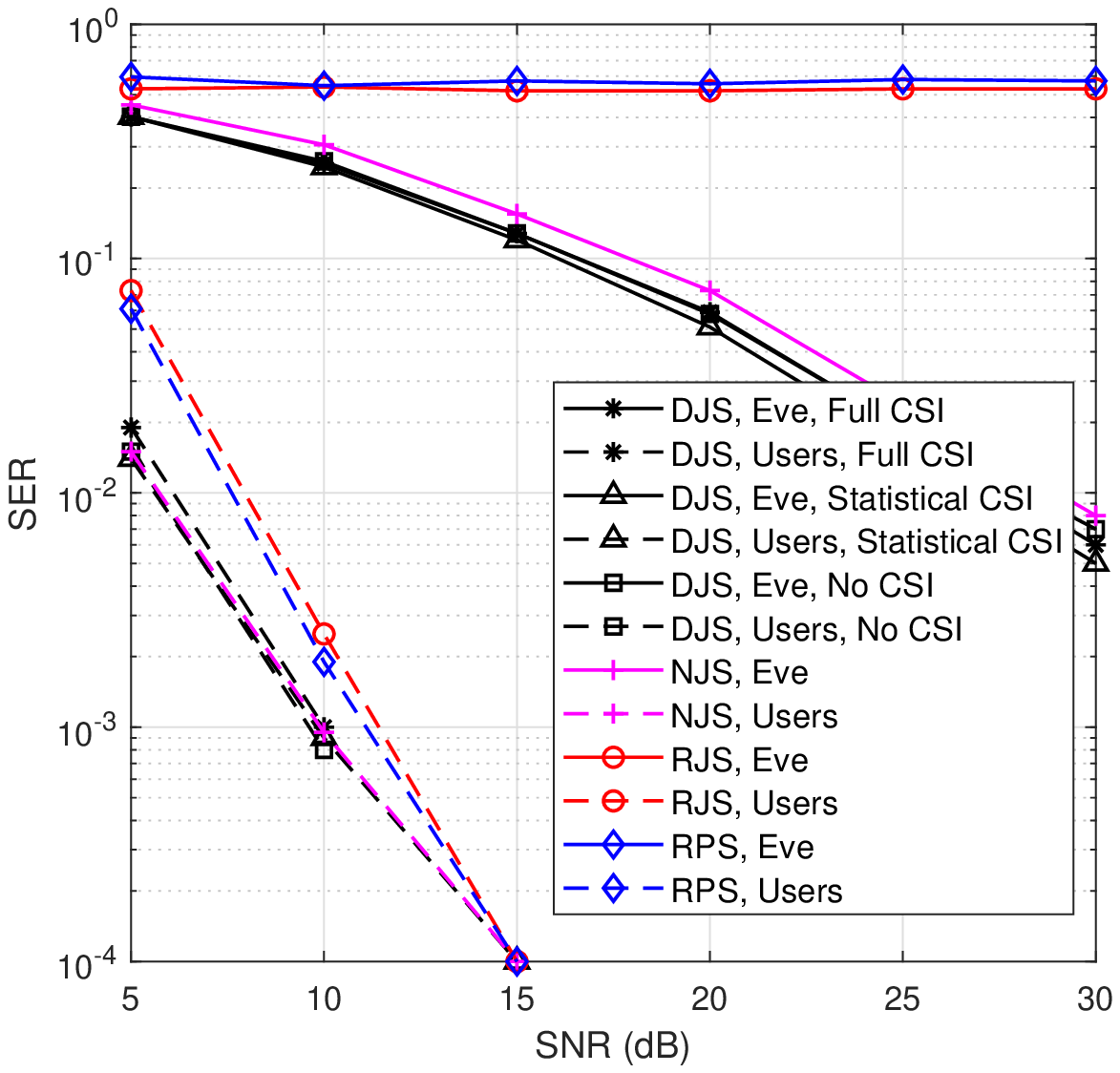}
\caption{QPSK: SER v.s. transmit SNR in a smart Eve case, $N=6,K=2$.}
\label{fig:side:b}
\end{minipage}
\vspace{-0.05 in}
\end{figure*}

First, we examine the transmit power of the proposed problem $\mathcal{P}_1$ and the DJS scheme in \cite{19} with QPSK modulation under the case with full Eve's CSI. As depicted in Figs. 4-5, the secrecy performance DJS is evaluated with the values $\Gamma_e=\{-5\;\text{dB}, 0\;\text{dB}, 5\;\text{dB}, 10\;\text{dB},0\}$, and the SNR at $\text{U}_k$ is assumed as $\Gamma_k=\Gamma$. Fig. 4 indicates that the proposed power minimization problem outperforms the DJS scheme in terms of the average transmit power, especially in the low required SNR regimes. In Fig. 5, the transmit power gain, which is defined as the difference between the power obtained from DJS and the proposed $\mathcal{P}_1$, clearly shows the advantage of the proposed scheme in power efficiency. Besides, we also study a special case of $\Gamma_e=0$, which represents that no signal is leaked out to the Eve. It is observed that the case of $\Gamma_e=0$ requires more transmit power than the optimal one, and indicates that such absolute security is at the expense of more transmit power compared with the proposed scheme.

Next, we present the SER performance of the problems $\mathcal{P}_2$-$\mathcal{P}_4$ along with the increasing maximum transmit SNR, i.e., $P_{s}/\sigma_k^2$, in Figs. 6-7. Each figure shows that when the transmitter knows full Eve's CSI, the SER performance at the Eve is better than those of the cases with statistical and no Eve's CSI since the wiretapped signal is designed in the destructive regions of the information symbol. However, the SER at the users is a little worse than those two cases due to the constricted feasible region. Besides, we also demonstrate that the proposed Algorithm 1 can achieve almost the same SER performance as CVX tool. Note that the simulation is performed on an Intel Core i7-6700 CPU 16 GB RAM computer with 3.4 GHz, Fig. 8 shows that the Algorithm 1 needs extremely less execution time than using CVX tool, which only occupies 6$\% \sim 8\%$ of the time of CVX, and improve the efficiency significantly.

\begin{figure}
\centering
\includegraphics[width=3.1 in]{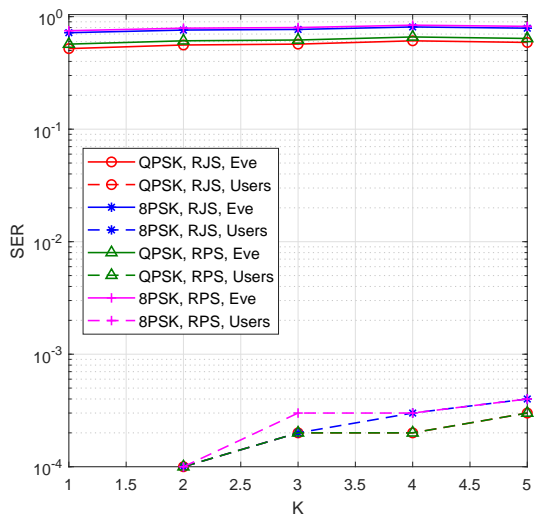}
\caption{SER v.s. $K$ users in RJS and RPS, $N=6$.}
\vspace{-0.1 in}
\end{figure}
Considering the case that the eavesdropper is smart, we first evaluate the constellation diagrams of the received signals at user 1 and Eve of the comparable schemes DJS and NJS in Fig. 9 over 1000 channel uses with SNR=15 dB. Obviously, we can indicate that the smart Eve can decode the information symbol at high probability by ML detection since it can formulate the corresponding signal to the information symbol. Moreover, we also display the constellation diagrams of the proposed schemes RJS and RPS with SNR=10 dB and $\rho=P_0/P_s=0.5$ in Fig. 10. It is obviously observed that the received symbols at $\text{U}_1$ locates in the expected constructive zone consisting with the information symbol, while the wiretapped symbols at Eve randomly distribute in all regions, so that the security can be guaranteed.

Moreover, we compare the SER performance of the proposed RJS and RPS with the DJS and RJS in Fig. 11 along with the increasing SNR, where the comparable schemes are operated by SINR balancing problem similarly as our proposed schemes for fairness, and $\rho=0.5$. It shows that DJS and NJS are hard to ensure the security of the information symbol when Eve uses ML detection, while our proposed schemes can protect the information signal successfully. The reason is that the jamming signal or the precoders in the comparable schemes are designed based on the CI conditions without randomness. While for the proposed RJS and RPS, we reserve the randomness of the jamming signal and the precoder ${\bf{p}}$ so that the wiretapped signal varies randomly in each channel use. Besides, we observe that RPS outperforms the RJS scheme at the users, which is due to the fact that the precoder ${\bf{p}}$ achieved by strict CI condition attributes to the decodability of the legitimate users.

Finally, we explore the SER performance along with the increasing number of the users under QPSK and 8PSK in Fig. 12, where SNR=15 dB and $\rho=0.5$. Obviously, the proposed RJS and RPS can protect the information symbols even with plenty of the eavesdroppers on the condition that $N\ge K+1$. In addition, the SER performance becomes worse when the number of the users increases, and RPS under both QPSK and 8PSK performs better than RJS due to the improved power efficiency at the transmitter in RPS.

\section{Conclusion}
This paper studied the physical layer security issues of a $K$-user MISO wiretap channel when there exists a common or a smart eavesdropper with SLP schemes. For the network with a common eavesdropper, we achieved additional transmit power savings of the network with full Eve's CSI by introducing the `complete destructive region' and jointly optimizing the threshold of the wiretap SINR. In order to improve the secrecy performance of the network, we analyzed the SINR-balancing problems with full, statistical, and no Eve's CSI, and optimized the SINR thresholds at both legitimate users and eavesdropper with the precoders. For the case with a smart eavesdropper with ML decoding strategy, we proposed the RJS and the RPS based on SLP strategy to protect the information symbols. We further proposed a simplified iterative algorithm to settle the convex optimization problems, and obtained a closed-form solution of the precoders. Taylor expansion has been utilized to transform the non-convex problems into convex ones. Simulation results showed that the proposed power-minimization problem outperforms the DJS in power efficiency, and the RJS and RPS perform much better than DJS and NJS in terms of the SER. Meanwhile, the proposed algorithm significantly improved the computation efficiency of the SLP problems.


%

%
%
%
%
%

\ifCLASSOPTIONcaptionsoff
  \newpage
\fi
\end{document}